\documentclass[12pt]{article}
\usepackage{mathtools}
\usepackage{amsmath}
\usepackage{amsfonts}
\usepackage{graphicx}
\usepackage{grffile}
\usepackage{helvet} 
\usepackage{relsize}
\usepackage{multirow}
\usepackage{tabularx}
\usepackage{amssymb}
\usepackage{parskip}
\usepackage{natbib} 
\usepackage{inputenc}
\usepackage{amsthm} 
\usepackage{comment}
\usepackage{adjustbox}
\usepackage[colorinlistoftodos]{todonotes}
\usepackage{lineno}
\usepackage{cleveref}
\usepackage{comment}
\usepackage{graphicx,psfrag,epsf}
\usepackage{enumerate}
\usepackage{natbib}
\usepackage{algorithm} 
\usepackage{algpseudocode} 
\usepackage{url} 
\usepackage{xcolor}
\usepackage[algo2e]{algorithm2e}
\usepackage{xr}
\usepackage{pdfpages}

\makeatletter
\newcommand*{\addFileDependency}[1]{
  \typeout{(#1)}
  \@addtofilelist{#1}
  \IfFileExists{#1}{}{\typeout{No file #1.}}
}
\makeatother

\newcommand*{\myexternaldocument}[1]{
    \externaldocument{#1}
    \addFileDependency{#1.tex}
    \addFileDependency{#1.aux}
}

\myexternaldocument{supplementary_material}

\newcommand{\blind}{0}

\newcommand{\bmc}{\hat{\beta}_{mc}}
\newcommand{\E}{\mathbb{E}}
\newcommand{\V}{\mathbb{V}}

\newtheorem{theorem}{Theorem}[section]

\newtheorem{remark}{Remark}

\algrenewcommand\algorithmicrequire{\textbf{Input:}}
\algrenewcommand\algorithmicensure{\textbf{Output:}}

\begin{document}

\def\spacingset#1{\renewcommand{\baselinestretch}%
{#1}\small\normalsize} \spacingset{1}


\if0\blind
{
  \title{\bf Inferring on joint associations from marginal associations and a reference sample}
  \author{Tzviel Frostig\thanks{
    The authors gratefully acknowledge \textit{the Israeli Science Foundation grant 2180/20, and the Planning and Budgeting Committee TAU Brain-MRI Bank Project grant	.}}\hspace{.2cm}\\
    Department of Statistics and Operation Research, Tel Aviv University\\
    and \\
    Ruth Heller \\
    Department of Statistics and Operation Research, Tel Aviv University}
  \maketitle
} \fi

\if1\blind
{
  \bigskip
  \bigskip
  \bigskip
  \begin{center}
    {\LARGE\bf Inferring on joint associations from marginal associations and a reference sample}
\end{center}
  \medskip
} \fi

\bigskip
\begin{abstract}

We present a method to infer on joint regression coefficients obtained from marginal regressions using a reference panel. 
This type of scenario is common in genetic fine-mapping, where the estimated marginal associations are reported in genomewide association studies (GWAS), and a reference panel is used for inference on the association in a joint regression model.
We show that ignoring the uncertainty due to the use of a reference panel instead of the original design matrix, can lead to a severe inflation of false discoveries and a lack of replicable findings.
We derive the asymptotic distribution of the estimated coefficients in the joint regression model, and show how it can be used to produce valid inference. We address two settings:  inference within  regions that are pre-selected, as well as within regions that are  selected based on the same data. By means of real data examples and simulations we demonstrate the usefulness  of our suggested methodology.


\end{abstract}

\noindent%
\vfill

\newpage
\spacingset{1.45} 
\section{Introduction}
    

        We consider the standard linear model: a quantitative response vector $y \in \mathbb{R}^n$ ($n$ being the number of observations), which is modeled as a linear function of $X$, 

	\begin{equation} \label{Eq:model}
	y = X\beta + \epsilon, \quad \epsilon \sim (0, \sigma^2 I_n),
	\end{equation}
    
    where $X$ is an $n \times p$ normalized covariates matrix and $p$ is the number of covariates. The ordinary least squares (OLS) estimate of the coefficients is $ \hat{\beta} = (X'X)^{-1} X' y$ for $p<n$ and $X$ of full rank. Conditioning on $X$, the distribution of $\hat{\beta}$ is 
	
	\begin{equation} \label{Eq:orgsim}
	\hat{\beta}|X \sim N_p\left(\beta, \sigma^2 \left(X' X\right)^{-1} \right).
	\end{equation}
		
	The distribution of $\hat{\beta}$ given $X$ is exact if $\epsilon$ follows a normal distribution, and it holds asymptotically (in $n$) otherwise. 
	
	We address the setting in which $n_o+n_r$ covariate vectors are sampled independently from the same distribution, but the analyst only has full access to the $n_r$ covariate vectors and to summary level statistics using the first $n_o$ covariate vectors and an outcome. Throughout, we  use $r$ and $o$ to indicate `reference' versus `original' data  (so $n_r$ refers to the number of observations in the reference panel).  
	We denote by $X_o$ and $X_r$ the covariate matrices using the $n_o$ and $n_r$ covariate vectors. The specific information  the analyst has access to is the marginal association of each of the $p$ covariates with the outcome, $X_o'y_o$, as well as the reference panel, $X_r$. \cite{yang2012conditional} suggested estimating the coefficients by plugging into the OLS estimate $X_r'X_r$ for $X'X$ and $X_o'y_o$ for $X'y$, see equation \eqref{Eq:bmcest}. 
    Their suggestion has been used in many subsequent studies, see e.g.,  \cite{horikoshi2016genome,van2012seventy, marouli2017rare} and more.

    This setting is standard in GWAS where the goal is to find genomic regions associated with a particular phenotype \citep{schaid2018genome}. The covariates are the single-nucleotide polymorphisms (SNPs) and the outcome is a phenotype or a trait of interest. 
	Since GWAS usually involves more SNPs than observations, each SNP association is typically estimated marginally without considering the other SNPs. The marginal associations can be large for non-causal SNPs since they are in linkage disequilibrium (LD) with a causal SNP \citep{schaid2018genome}. The process of extracting candidate `causal' SNPs (where a SNP is 'causal' if it is associated with the phenotype after conditioning on all other SNPs) from GWAS is called fine-mapping.
	
	In order to extract causal SNPs from GWAS, the SNPs are first divided into independent regions of interest according to the statistical significance of their marginal association with the phenotype and their LD.
	To determine the causal SNPs within a region, it is possible to use penalized regression approaches such as LASSO and elastic-net \citep{tibshirani1996regression, zou2005regularization, vignal2011using, waldmann2013evaluation} which are applicable even when the number of observations is smaller than the number of SNPs in the region. In such case, the  candidate SNPs of interest are  the subset of SNPs where the effect size is not shrunken to zero, and inference on the associated SNPs, while controlling a relevant error rate, has to  carried out by accounting for the selection step.  See \cite{lee2016exact, taylor2018post,  loftus2015selective} for such adjustments. 
	
	A different approach is to further divide the regions into smaller regions in which the number of SNPs considered in each such region is smaller than the number of observations \citep{schaid2018genome}. The advantage over the penalized reggression approaches is that  standard regression methods can be applied and the inference is more straightforward (i.e., $p$-values and confidence intervals can be easily computed as long as the full data is available). Moreover, even though the inference within smaller regions is marginalised over the SNPs outside these regions, it is still preferred over complete marginalization, where the phenotype is regressed on each SNP separately. This is the approach we consider in this paper.   
	
	There are also Bayesian approaches specifically designed for fine-mapping, such as FINEMAP \citep{Benner16} and RSS \citep{zhu2017bayesian}. In the Bayesian methods, all possible combinations of SNPs are considered (2 to the power of number of SNPs) in order to find the posterior inclusion probability of each SNP. They are usually limited by the number of causal SNPs considered. Moreover, they rely on a prior distribution on the number of causal SNPs, which can bias the inference. For a review of fine-mapping, see \cite{schaid2018genome}.

     Although sharing $X_o$ between analysts is useful for fine-mapping, it can be problematic due to: (i) logistical issues since the data is large (ii) privacy issues that come with using such sensitive data \citep{zhu2017bayesian}.  So, it should not come as a surprise that GWAS researchers usually do not disclose their study's genetic data but only the marginal regression results. Fine-mapping may still be possible using public genetic repositories, such as the 1000 Genome Project, HapMap and the UK-biobank \citep{10002012integrated, international2003international, UKBiobank}. The genetic data in them is used to estimate the LD matrix and plugged in to transform the estimated marginal associations into joint associations  \citep{yang2012conditional}. The resulting estimator, displayed in Eq. \eqref{Eq:bmcest}, is  a consistent estimator of $\beta$ when $n_r, n_o\rightarrow \infty$. But, it has extra variance due to the use of a reference panel (rather than the original set of covariates) which relies on the ratio of $\frac{n_r}{n_o}$ and not on the total number of samples. Failing to account for this in the fine mapping inference can lead to an inflated type I errors and non-replicable findings. 
     
	 \cite{benner2017prospects} first discussed the problem of using reference panels in the context of Bayesian fine-mapping and suggested storing larger reference panels. However, in Frequentist estimation, the problem persists for any size of reference panel, as shown in in Theorem \ref{theorem:general_dist}. 
    
	Our goal is to use a reference panel with the marginal regressions obtained from a single study to infer on the parameters, taking into account the variance inflation due to the use of the reference panel.
	We address two settings. First, the region of interest is set in advance. Second, it is selected based on the data, thus requiring adjusting for selection bias. Methods that do not require access to $X_o$ for region selection condition on the selection event, e.g.,  by and Tag-SNP \citep{lee2014exact}, or using an aggregation test \citep{Heller19}. 	
	To correct for the selection bias, we incorporate the Post Selection after Aggregate Tests (PSAT, \cite{Heller19}) framework, allowing for valid inference after selection. 

	
    We show how estimating the covariance matrix based on the reference panel increases the coefficient estimators' variance in both settings. To account for the variance inflation, we rely on results regarding the correlation matrix variance under general conditions \citep{browne1986asymptotic, neudecker1990asymptotic}. We obtain the asymptotic distribution of the estimated joint coefficients and suggest an estimator of the additional variance. 
	
	The rest of the paper is organized as follows. In \S 2 we formalize the problem and goal. In \S 3, we present our main theoretical result regarding inference on joint effects without selection. In \S 4, we present the PSAT framework using marginal associations and a reference panel. In \S 5 we present numerical results. In \S 6 we analyze BMI data from the UK-biobank,  and the Dallas Heart Study data. The latter is an example of using PSAT with our suggested method. We conclude with a discussion in  \S 7. 
	The supplementary material (SM) contains the proofs, and further details regarding the suggested method's computational complexity and implementation. 
    
    \section{Set-up and goal} \label{sect:setup}
     We consider the setting for a pre-specified single region where the number of observations is larger than the number of covariates. In \S \ref{sect:PSAT} we extend the suggested method to multiple regions that we first screen, and only regions judged to contain signal are further analyzed. 
    
    Let  
    $X_r, X_o$ and $y_o$ denote the standardized  columns (i.e., each column has mean  0 and standard deviation  1).  Let $X_{i,.}$ refer to the covariates of an observation indexed by $i$.  
    Let $X^*_o$ and $X^*_r$ denote the centered covariate matrices, so $\E(X^*_i) = 0$ and $\V(X^*_{i, .}) = \Sigma$, where $\E$ and $\V$ are the expected value and variance operators. 
    The marginal associations  are
    $$ \hat{\beta}^m_o =  \left( d(X_o'X_o) \right)^{-1} X_o' y_o = n_o^{-1} X_o'y_o,$$
    where $d: \mathbb{R}^{p} \times \mathbb{R}^{p} \rightarrow \mathbb{R}^{p} \times \mathbb{R}^{p}  $ is the matrix operator that sets all non-diagonal elements of a square matrix to zero. The second equality follows since $X_o$ is standardized.

    Our available data is the reference panel and the marginal associations from the original study, $ \hat{\beta}^m_o$. Our goal is fine-mapping: to identify the SNPs  associated with the phenotype in a region of interest.
    
    The conditional distribution of the marginal associations is, 
	
	\begin{equation} \label{eq:marginal}
	\hat{\beta}_o^m | X_o \sim N_p \left( \hat{R}_o \beta, \frac{\sigma^2}{n_o} \hat{R}_o \right). 
	\end{equation} 
	
	where $ \hat{R}_{o} = \frac{X_o' X_o}{n_o}$ is a consistent estimator for $R$, the correlation matrix of $X$. $\hat{R}_o$ (or $X_o$) is not available to the analyst. Instead, we can use  $\hat{R}_r = \frac{X_r'X_r}{n_r}$, which is also a consistent estimator of $R$. \cite{yang2012conditional} suggested estimating $\beta$ by 
	
	\begin{equation} \label{Eq:bmcest} 
	\hat{\beta}_{mc} = \frac{n_r}{n_o} (X_r'X_r)^{-1} X_o' y_o = \hat{R}_r^{-1} \hat{\beta}_o^m.
	\end{equation} 
	
	The  distribution of $\bmc$ is approximated by $N_p(\beta,  n_o^{-1} \sigma^2 \hat{R}_r^{-1})$  in \cite{yang2012conditional}, but this approximation ignores the fact that $X_o$ and $X_r$ are not identical.
	We show in Theorem \ref{theorem:general_dist} that even when $X^*_r$ and  $X^*_o$ are sampled independently from the same population, the inference is biased because $X^*_r \neq X^*_o$. 
	
	Using the above approximate distribution of $\bmc$ for inference is henceforth addressed as the  \textit{naive} approach. It leads to an under-estimation of the $\hat{\beta}_{mc}$ variance and an increase in false positives (FP). This causes a lack of replicability for studies based on summary-level statistics. For example, the FP are almost $50\%$ of the discoveries in  some of the settings in Fig. \ref{Fig:fine_fdr} A. 	For valid inference on $\beta$, the additional variance due to the use of the reference panel needs to be accounted for. Our goal is to quantify this variance and suggest a valid testing procedure. 
	
	\section{Inferring the joint effects without selection} \label{sect:no_select_infer}
	
    In order to infer on $\beta$ we derive the asymptotic distribution of $\bmc$, given in Theorem \ref{theorem:general_dist}. The theorem relies on obtaining the covariance matrix of the vectorized correlation matrix as derived by \cite{neudecker1990asymptotic}, 
    see their result reproduced in Theorem \ref{Thm:Neudecker}. 
    
	\begin{theorem} \label{theorem:general_dist}
		Assume the following for $X_{i,.} \in \mathbb{R}^p$: $\V(X_{i,.}) = R$ is of full rank;  a bounded fourth moment, $\forall j \; \E(X_{i,j}^4) < \infty$. Assume also that $X_o, X_r$ are the matrices with standardized columns of  the independent identically distributed rows $X_{i,.}, i=1, \ldots, n_o+n_r$. 
		The asymptotic distribution of $\hat{\beta}_{mc}$  as $\lim_{n_r,n_o \rightarrow \infty}\frac{n_r}{n_o} \rightarrow c \geq 0 $ is 
		
	\begin{equation} \label{eq:thm}
		\lim_{n_r,n_o \rightarrow \infty} \sqrt{n_r}  (\hat{\beta}_{mc} - \beta) \sim N_p(0,  c\sigma^2 R^{-1} +  (1 + c)(\beta' \otimes R^{-1}) \boldsymbol{V}_R (\beta \otimes R^{-1})), 
	\end{equation}
	
	where  $\otimes$ is the Kronecker product.
	\end{theorem}
	See SM \S \ref{sect:proof} for the proof.
	Let  $ \Sigma_{mc} = \V(\hat{\beta}_{mc})$, and  $\hat{\Sigma}_{mc}$ its estimator. Examination of the asymptotic covariance provides the following insights. First, if $\beta = 0$, then using the reference panel does not increase the asymptotic variance of the estimator, implying that the naive approximation of $\bmc$ is correct under the global-null hypothesis of no association between all $p$ variables and the response. 
    Second, the variance expression contradicts the claim in \cite{benner2017prospects}, that large reference panels can eliminate the problem. 
    The variance inflation depends on the ratio $\frac{n_r}{n_o}$, so even while $n_r, n_o \rightarrow \infty$ there can be substantial variance inflation.
    
    \begin{remark} 
    Using the \textit{naive} variance estimator will raise the number of false discoveries as $\frac{n_o}{n_r}$ increase, even though $\V \left( \hat{\beta}_{mc} \right)$ decrease (if overall $n_o + n_r$ increase). According to Theorem \ref{eq:thm}, $\V(\hat{\beta}_{mc}) = O(n_r^{-1})$, while the naive estimate of the variance is $O(n_o^{-1})$. Thus, as $n_o$ increase $\hat{\Sigma}_{mc}$ is underestimated, leading to more false discoveries. 
    Even when $c=0$, there is still variance inflation, although in the variance is still in the same magnitude as in the naive estimate, $O(n_o^{-1})$. 
    \end{remark}
    

    \subsection{Estimating $\Sigma_{mc}$} 
	
	In order to estimate $\Sigma_{mc}$, the parameters are replaced by their respective estimators. So, $R$ is replaced by $\hat{R}_r$, and $c$ is replaced by $\frac{n_r}{n_o}$. We proceed to describe the methods to estimate the remaining parameters: $\boldsymbol{V}_R$, $\beta$, and $\sigma^2$. 
	
	\subsubsection{Estimation of $\mathbf{V}_R$}  \label{sect:estvr}
    
     $\mathbf{V}_R$ is a function of the variance of the vectorized covariance matrix $\mathbf{V}_{\hat{\Sigma}} = \V\left(vec(\hat{\Sigma}) \right)$, where $vec: \mathbb{R}^{p \times p } \rightarrow \mathbb{R}^{p^2}$ is the vectorization operator, which concatenates the $p$ columns of the square matrix to a vector. Therefore, $\mathbf{V}_{\hat{\Sigma}}$ needs to be estimated. 
     
     If $X^*_{i,.}$ follows a Gaussian distribution $\mathbf{V}_{\Sigma}$ can be estimated as $ \hat{\mathbf{V}}_{\hat{\Sigma}_r} = 2 M_s (\hat{\Sigma}_r \otimes \hat{\Sigma}_r)$, where $\hat{\Sigma}_r =\frac{(X^*_r)' (X^*_r)}{n_r}$ (as $X_r^*$ is centered) and $M_s$ is defined in Theorem \ref{Thm:Neudecker}. In the case where the $X^*_{i,.}$ does not follow a Gaussian distribution the parameter, $\mathbf{V}_{\hat{\Sigma}_r}$, is estimated empirically. Denote the estimated variance based on observation $i$ by $\hat{\Sigma}_{r_i} = (X^*_{r_{i,.}})(X^*_{r_{i,.}})'$ . The variance of the covariance matrix is estimated by $\hat{\mathbf{V}}_{\hat{\Sigma}_r} = n_r^{-1} \sum_{i=1}^{n_r} \left(vec(\hat{\Sigma}_{r_i}) - vec(\hat{\Sigma}_{r}) \right) \left(vec(\hat{\Sigma}_{r_i}) - vec(\hat{\Sigma}_{r})\right)'$.
	
	Estimating $\mathbf{V}_{\Sigma}$ involves computing the outer-product of $n$ vectors of length $p^2$, resulting in a complexity of $O(np^4)$.Estimating $\mathbf{V}_{\Sigma}$ for covariates from the multivariate normal distribution is only $O(p^4)$, see \S \ref{sect:computevar} for further details.  
	
	\subsubsection{Estimation of $\sigma^2$ in $\Sigma_{mc}$}

	Since $y$ is standardized to have variance one, a conservative estimate is $\hat{\sigma}^2 = 1$.
	For small $\sigma^2$ this would result in a loss of power. However, since 	
	\begin{equation}
    \sigma^2 = 1 - var(X\beta) = 1 - \beta' R \beta,
	\end{equation}
	
	a natural estimator is
	\begin{equation} \label{eq:var_est}
	    s^2 = 1 - \hat{\beta}_{mc} \hat{R}_r \hat{\beta}_{mc}. 
	\end{equation}
	
     Unless stated otherwise, we use the latter estimate in our simulations and data analyses.

	\subsubsection{Estimation of $\beta$ in $\Sigma_{mc}$} \label{sect:betathres}

	The vector of coefficients of interest, $\beta$, also appears in the expression for $\Sigma_{mc}$. Of course, $\beta$ is unknown since it is the target for inference. 	The natural candidate is the point estimate, $\bmc$. However, in order to ease the computational burden, we suggest a threshold version of $\bmc$, based on the assumption that some entries of $\beta$ are zero and their corresponding entries in $\bmc$ are small and contribute little to $\hat{\Sigma}_{mc}$. 
	
	To obtain the threshold estimator, each coefficient is tested, $H_{0,i}: \beta_i = 0$. The test statistic using the naive variance estimator of $\bmc$ is $ T^* =   \frac{\sqrt{n_o}}{\hat{\sigma}} d(\hat{R}_r)^{-\frac{1}{2}} \bmc . $ 
	The resulting test is not valid since the variance is underestimated, making the test too liberal. Here lies the crux of the heuristic, how to select the threshold?  
    We suggest using a threshold corresponding to the significance level employed in the principal analysis. It ensures that any entry of $\beta$ which has not passed the threshold would not be discovered in the principal analysis (since $d(\hat{\Sigma}_{mc}) \geq d(R^{-1}_{r})$, the inequality refers to each entry of the matrices).  
    
    The main drawback of the heuristic is in the case where most entries of $\beta$ are non-zero, and using the threshold version of $\bmc$ leads to an underestimate of $\Sigma_{mc}$ for $\beta_i = 0$ and an inflated type I error probability. However, in our simulations, we found that when there are up to $0.25$ of non-zero entries of $\bmc$, the multiple testing procedure still maintains the error rate at the nominal level.

	\section{Inferring the joint effects following selection}  \label{sect:PSAT} 
	
	The fine-mapping goal is to identify SNPs that are significantly correlated with a specific phenotype. Often, to accomplish this goal, the genome is separated into regions, and SNPs are only examined in regions deemed interesting. A typical example is selecting regions according to the tag-SNP coefficient (if it is above a certain threshold).  
	After the regions are selected, the goal is to identify the specific SNPs associated with the phenotype in a joint analysis. If the selection and follow-up identification are carried out on the same data, the selection process needs to be accounted for in order for the inference to be valid.	

    The approach of adjusting for selection by conditioning on the selection event has been studied extensively in recent years \citep{taylor2014exact, loftus2015selective, lee2016exact, reid2018general}. 
	\cite{Heller19} suggested a method for inferring on individual parameters (SNPs in our scenario), following the rejection of the global-null hypothesis with a linear or quadratic aggregate test.

	
	Next, we review the post selection after aggregate test (PSAT) method and show how to adjust for selection when only $X_r$ and $X_o'y$ are known.
	

	\subsection{Post selection inference after aggregate tests (PSAT)} \label{sect:ReviewPSAT}

	A typical fine-mapping strategy is to select regions for further analysis if their tag-SNP absolute marginal coefficient, $|\hat{\beta}^m_o|$ is above a certain threshold. Since in the suggested use-case the original reference is not available, the selection must be taken into consideration while using only the reference sample.  First, the selection event is described in terms of $\bmc$ and $\hat{R}_r$, 
	
	\begin{equation}
	    S(\bmc, \hat{R}_r) = \hat{\beta}_{mc}' \hat{R}_r e^{*} e^{*'}  \hat{R}_r \hat{\beta}_{mc}  > t,
	\end{equation}
	
	where $e^{*}$ a vector of length $p$ with entry 1 in the respective tag-SNP index and 0 elsewhere. At this point, the PSAT method can be applied to infer on any linear combination of the coefficients, $\eta \bmc$. PSAT requires additional conditioning on $W = (I - c \eta')\hat{\beta}$, where $c= (\eta' \Sigma_{mc} \eta)^{-1}\Sigma_{mc}^{-1}\eta$. The resulting distribution is 
	
	\begin{equation} \label{eq:psat}
		\eta' \bmc|S>t, W, X \sim TN(\eta' \beta, \eta' \Sigma_{mc} \eta, \mathcal{A}(W)),
	\end{equation}

	where $TN(a,b,d)$ denotes the truncated normal with expected value of $a$, variance $b$ and truncation region of $d$. The truncation region, $\mathcal{A}(W)$, is given in Lemma A.1 of    of \cite{Heller19}. The PSAT procedure can also be applied to linear based selections, an implementation of the method can be found at \url{github.com/tfrostig/PSAT}. 
	
	\begin{remark}
	When the global-null hypothesis holds, $\beta_i = 0 \; \forall i$, then the naive approximation of $\hat{\beta}_{mc}$ distribution is correct. 
	Therefore, when applying PSAT in using summary level statistics and a reference panel, the computationally expensive estimate of $\Sigma_{mc}$ is only computed in regions where the aggregate test was rejected. 
	\end{remark}

	\subsection{Estimating $\Sigma_{mc}$ after selection} \label{sect:sig_after_selection}
	
	Following selection, we also need to adjust our procedure for estimating $\Sigma_{mc}$. Estimation of $V_R$ does not change, but the estimates of $\sigma^2$ and $\beta$ do. 
	
	Due to selection, $|\bmc|$ is biased upwards, and this could lead to an upward bias in the estimation of $\Sigma_{mc}$. 
	
	According to Eq. \eqref{eq:psat}, the linear combination follows a truncated normal distribution, so a conditional maximum likelihood estimate (MLE), $\tilde{\beta}_{mc}$ can be computed,
	
	\begin{equation} \label{eq:cond_beta} 
	\tilde{\beta}_{mc} = \arg \max_{\beta} \{l
	(\beta) - \log P(S > t) \}
    \end{equation}
	
	where $l$ is the log-liklihood for $\beta$, see \cite{Heller19}, \S 4.1 for further details.
	
	The MLE is unbiased after selection, thus replacing $\bmc$ with $\tilde{\beta}_{mc}$ in the estimate of $\hat{\Sigma}_{mc}$  and $\hat{\sigma}^2$ reduces their bias. To threshold the $\tilde{\beta}_{mc}$, we apply the same process as described in \S \ref{sect:betathres}, but estimate $\sigma^2$ in Eq. \ref{eq:var_est}, by replacing $\beta$ with $\tilde{\beta}_{mc}$.  
	
	
	 \begin{algorithm}[H]
    \small\caption{PSAT analysis using reference panel (without thresholding)}
    \label{alg:test} 
    \SetKwInOut{Input}{Input}\SetKwInOut{Output}{Output}
    \Input{$\hat{\beta}^o_m, \hat{R}_r, \hat{R}_o, n_r,n_o, S$ } 
    \Output{$\mathbf{PV}$}
    \SetAlgoLined
    $\bmc = \hat{R}_r^{-1} \hat{\beta}_o^m$ \;
    
    $\hat{\sigma}^2 = 1 - \tilde{\beta}_{mc}' \hat{R}_r \tilde{\beta}_{mc}$ \; 
    
    \eIf{$S(\bmc, \hat{R}_r) < t$}{
    $\mathbf{PV}$
    $\boldsymbol{Return} \; $
    } 
    {
        Apply PSAT using $\bmc, \hat{R}_r, S, t$ , obtaining the conditional MLE of $\beta$, denoted by $\tilde{\beta}_{mc}$  (Eq. \ref{eq:cond_beta})
        
	    
	    Estimate $\boldsymbol{V}_R$ and $\sigma^2$ using $\tilde{\beta}_{mc}$, obtaining $\hat{\boldsymbol{V}}_R$ and $\tilde{\sigma}^2$ (\S \ref{sect:estvr} and Eq. \ref{eq:var_est})
	    
	    Let $\hat{\Sigma}_{mc} = \frac{n_r}{n_o} \tilde{\sigma}^2 \hat{R}_r^{-1} +  \frac{n_o + n_r}{n_o n_r}(\tilde{\beta}_{mc}' \otimes \hat{R}_r^{-1}) \hat{\boldsymbol{V}}_R (\tilde{\beta}_{mc} \otimes \hat{R}_r^{-1})$
	    
	    Obtain the conditional p-values, $\mathbf{PV}$, according to Eq. \eqref{eq:psat} and estimates  $\tilde{\beta}_{mc}, \hat{\Sigma}_{mc}, S, t$
	    
	    $\boldsymbol{Return} \; \mathbf{PV}$
	   
    }
    \end{algorithm}

	\section{Simulation studies} \label{sect:sim}
	We use simulations in order to assess the performance of our variability corrected approach using summary level data only. We have the following goals: (1) to assess the power loss incurred from using only summary level data and a reference panel; (2) to assess the validity of the approach for finite samples; (3) to compare the most generally applicable variability correction, which is computationally demanding, with the variability correction method which is theoretically justified only for Gaussian covariates. 
	
	We compare the performance of four methods to infer on $\beta$: (1) \textit{Full} - $X_o, y_o$ are given and the standard OLS approach is employed. The variance estimate therefore is $\frac{\hat{\sigma}^2}{n_o} \hat{R}_o^{-1}$. (2) \textit{Naive} method - does not take into account the additional variance caused by $X_r$. The variance estimator is $\frac{\hat{\sigma}^2}{n_o} \hat{R}_r^{-1}$. (3) \textit{Variability corrected (empirical)} - based on Theorem $\ref{theorem:general_dist}$, where $var\left( vec (\hat{\Sigma}_r) \right)$ is estimated empirically. (4) \textit{Variability corrected (Gaussian)} - $X_r, X_o$ are assumed to follow a Gaussian distribution and to estimate the variance of the sample covariance matrix we assume $var \left( vec \left( \hat{\Sigma_r} \right) \right) = 2 M_s (\Sigma \otimes \Sigma)$, $M_s$ is defined in Theorem \ref{Thm:Neudecker}.

	
	Within a simulated region, we denote by $s$ the set of significant SNPs, and by $s^*$ the set of  SNPs with effect. The false discovery proportion is $FDP = \frac{|s \setminus s^*|}{|s|}$, and the TPP is $\frac{|s \cap s^*|}{|s^*|}$, the average of both serves as an estimate of the FDR and power respectively.  In all simulations the Benjamini-Hochberg (BH) method \citep{benjamini1995controlling} is used at level 0.05. Unless stated otherwise, the number of repetitions in each simulation is 1000.  
    
    The data generation begin with sampling from $N(0, \Sigma)$, where $\Sigma_{i,j} = \rho^{|i-j|}$, in order to generate correlation between adjacent SNPs. Later the data is scaled or transformed into SNPs, $X_o, X_r$. The number of features is 20 unless stated otherwise.  
    
    The coefficients, $\beta$, are 1 where indices match $s^*$ and 0 elsewhere. The explained variance, or the heritability of the phenotype is denoted by, 
    
    \begin{equation} \label{eq:h}
        h = \frac{\beta' R \beta}{\sigma^2_\epsilon + \beta' R \beta}.
    \end{equation}

    We set $h$ by adjusting $\sigma^2_\epsilon$.
    
    Briefly, our key findings are as follows. First, that as expected the \textit{Naive} estimator fails to maintain the nominal FDR level. The problem exacerbates when $\frac{n_r}{n_o}$ is large, when the ratio is small, and both $n_r$ and $n_o$ increase the FDR approaches the expected level. Second, that the \textit{Variability corrected (Empirical)} is valid (i.e., provides an FDR level below the nominal level) in all scenario, while \textit{Variability corrected (Gaussian)} is only valid when indeed the distribution of $X$ is Gaussian. Third, that when the signal is denser (i.e., there are several SNPs/features with non-zero coefficients), the power of the method \textit{Variability corrected (Empirical)} decreases. Fourth, that the power loss of the \textit{Variability corrected (Empirical)} in comparison with the \textit{Oracle} method is not negligible, but the loss decreases as $n_r$ increase.

	

	\subsection{Gaussian distributed covariates} \label{sect:gauss_sim}
	
    $X_o^*, X_r^*$ are sampled from $N(0, \Sigma)$, where $\Sigma_{i,j} = 0.8^{|i-j|}$, and are then standardized in order  to obtain $X_o, X_r$. 
    The number of observations in the original study and in the reference panel vary: $n_o \in \{10000, 20000, 100000, 200000\}$;  $n_r \in \{500, 1000, 5000, 10000\}$. We consider three values of $h \in \{0.005, 0.01, 0.05\}$. 

	\begin{figure}[hbt!] 
		\centering
		\includegraphics[width=0.65\textwidth]{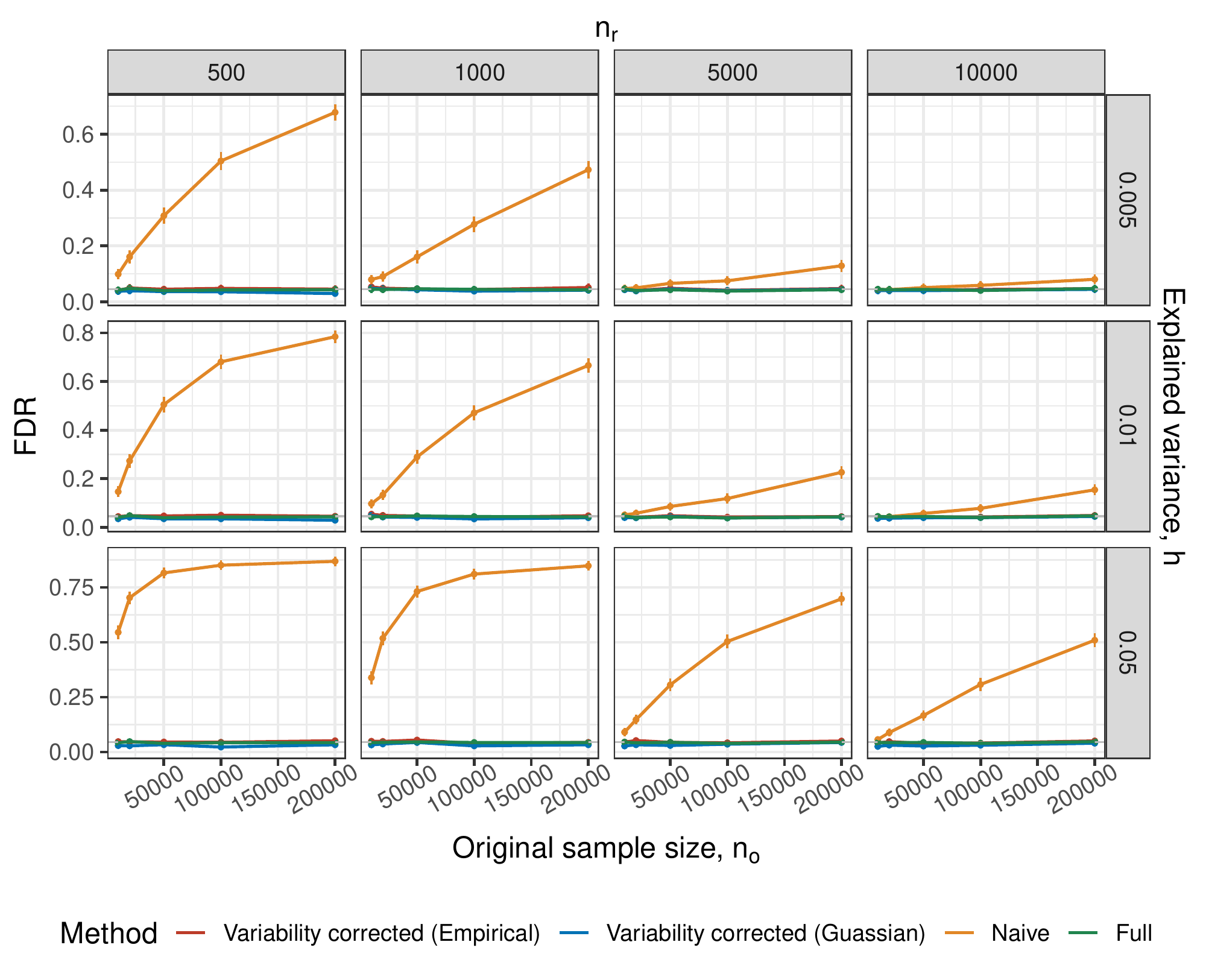}
		\caption{The FDR versus $n_o$ in settings with simulated Gaussian covariates. Each plot in the facet is a combination of the reference sample size $n_r$ (columns), and explained variance $h$ (rows, Eq. \ref{eq:h}). The  covariates with non-zero coefficients are $s^* = \{1, 20\}$,   and $\rho = 0.8$, across all scenarios. Vertical lines represent the 2 standard error around the estimate.   
		}
		\label{Fig:fine_fdr_gauss}
	\end{figure}
	
    
    In Figure \ref{Fig:fine_fdr_gauss}, the \textit{Naive} method fails to maintain the expected FDR level of $0.05$ in all settings considered. Even when considering a large number of observations in the reference sample ($n_r \geq 5000$), if the number of observations in the original study is sufficiently large, the \textit{Naive} method will fail. When the explained variance is small, it implies that $\beta$ is also small, leading to a smaller inflation, and thus the \textit{Naive} method is closer to the nominal FDR level. Both \textit{Variability corrected (Empirical)} and \textit{Variability corrected (Gaussian)} maintain the FDR across all scenarios. 
    
    
	\subsection{A setting mimicking genomic covariates} \label{sect:fabsim}
	
	We begin by generating $W \in \mathbb{R}^{(n_o+n_r) \times p}$ where $W_i \sim N(0, \Sigma)$. To generate the genotype matrices, $X_r^*, X_o^*$, we transform $W$ as follows. We apply  $g(. ; q): \mathbb{R} \rightarrow \{0, 1, 2\}$ to each coordinate of $W$, where $q$ is the minor allele frequency (MAF). The MAF is sampled from $Beta(1,2) $, then divided by two and clumped at $0.05$. The function is 
	
	\begin{equation}
		g(w, q) = \begin{cases} 
			0 , \quad w \leq Z_{1 - \frac{2}{3} q} \\
			1 , \quad Z_{1 - \frac{2}{3} q} <  w < Z_{1 - \frac{1}{3} q} \\ 
			2 , \quad w \geq Z_{1 - \frac{1}{3} q}
			\end{cases},
	\end{equation}
	
	where $Z_q$ is the $q$-'th quantile of the standard normal distribution. 


	
	We present the results for $s^* \in \{ \{1, 20\}, \{1,5,10,15,20\} \}$,  $\rho = 0.95$, $n_o = 10000$ and $h \in \{5 \times 10^{-4}, 0.0025, 0.01, 0.05\}$. We experimented with other parameters, but they were omitted for brevity, since the qualitative conclusions were similar. 
		
	\begin{figure}[hbt!] 
		\centering
		\includegraphics[width=0.685\textheight]{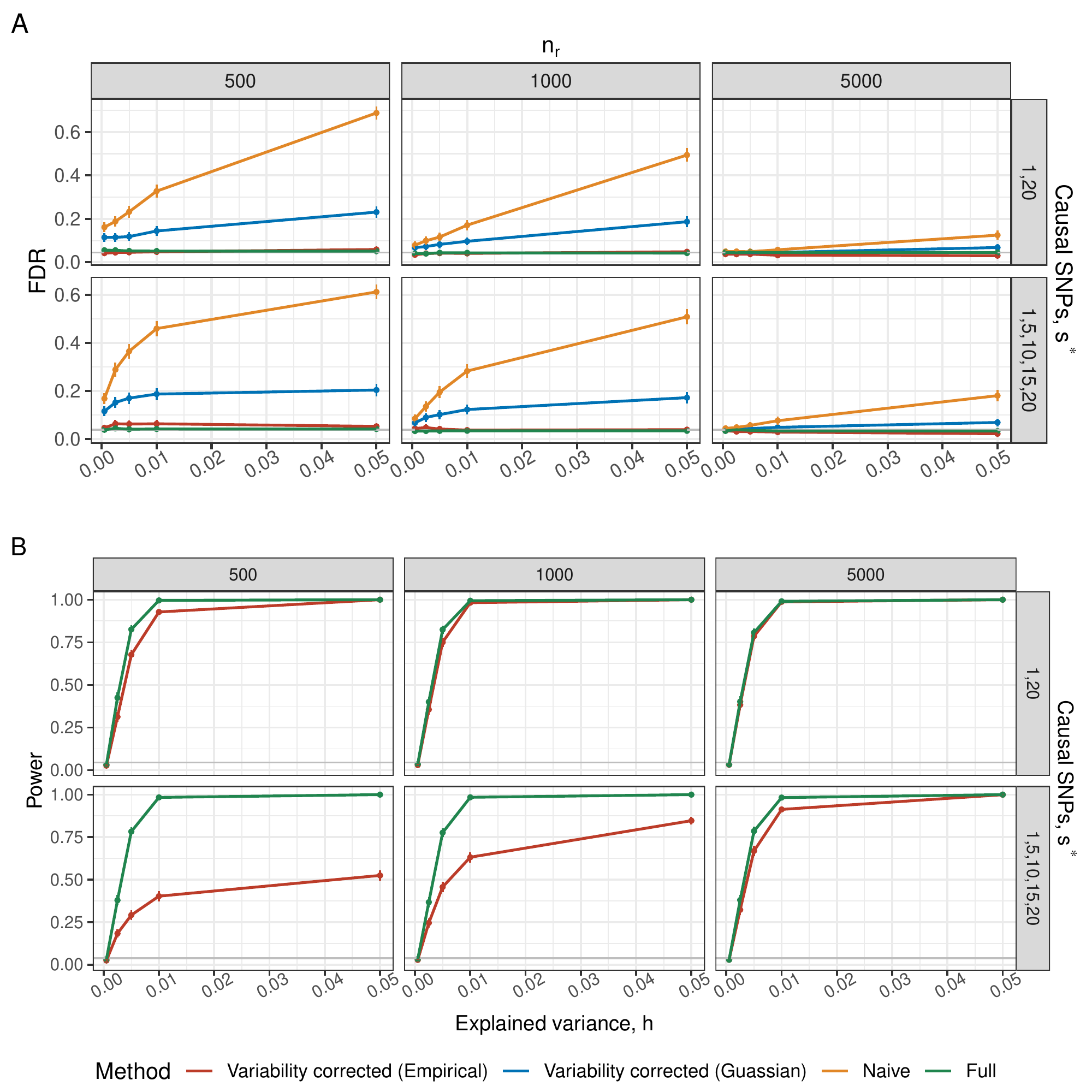}
		\caption{The FDR (A) and power (B) versus $h$ (Eq. \ref{eq:h}) in settings mimicking genomic covariates. Each plot in the facet is a combination of causal SNPs indices, $s^*$ (rows), and the reference sample size, $n_r$ (columns). The number of observations in the original study $n_o =10000$ and $\rho = 0.95$ are kept constant across all scenarios. Vertical lines represent the 2 standard error around the estimate.} 
		\label{Fig:fine_fdr}
	\end{figure}

    In Figure \ref{Fig:fine_fdr}A, the \textit{Variability corrected (Empirical)} method maintains the expected FDR across all scenarios and for all $n_r$. The \textit{Naive} method fails to maintain the FDR in all scenarios, even when $n_r =5000$. Furthermore, as $h$ increase the coefficients magnitude compared to $\sigma_\epsilon^2$ increase, leading to more unaccounted variance resulting in an increase in FDR of the \textit{Naive} method. 
	We can also see that the \textit{Variability corrected (Gaussian)} FDR decreases as $n_r$ increases. However, it fails to maintain the FDR in all scenarios. Therefore, the \textit{Naive} and \textit{Variability corrected (Gaussian)} methods are omitted from the subsequent power analysis. 

	In Figure \ref{Fig:fine_fdr}B, we can see the power across various data generation settings. A challenging scenario for the \textit{Variability corrected (Empirical)} method is $s^* = \{1,5,10,15,20\}$, due to all non-causal SNPs having a nonzero marginal association with at least one causal SNP. In this scenario lies the largest difference in power between the \textit{Full} and the \textit{Variability corrected (Empirical)} methods when $n_r =500$. 

	The power curves are very similar in the realistic scenario where only two SNPs are causal. When $n_r$ is sufficiently large, the power curves merge. This happens because the additional variance shrinks as $n_r$ increases (see Eq. \eqref{eq:thm}). For sparse signal, $s^* = \{1, 20\}$, $n_r = 500$ is sufficient even for strong correlation. For denser signal, $n_r$ has to increase for adequate power. 
	
	\subsubsection{The PSAT analysis pipeline}
	
    In many applications of fine-mapping, a Tag-SNP of a region is used to decide whether the region is worth further investigation or not. To mimic this scenario, we employ the same simulation strategy as detailed in the beginning in the of \S \ref{sect:fabsim}, but add a Tag-SNP located in index 10. 
	
	The region is selected if the Tag-SNP estimated marginal coefficient is above the threshold of $n_o \times Z_{1 - 0.05 / 20000}$ (corresponding to testing the marginal coefficient if there is no signal nor LD at $\alpha = 0.05$ and applying the Bonferroni correction for $20,000$ hypotheses). It imitates a GWAS where $20,000$ loci are being tested, each by their Tag-SNP using a Bonferroni correction. 	At each iteration, the noise, $\epsilon$, is re-sampled until the selection event occurs. All other simulation configurations are identical to those detailed in \S \ref{sect:fabsim}.

	Due to the incorporation of the PSAT procedure, we add another method, \textit{Variability corrected (MLE)}. This method differs from the \textit{Variability corrected (Empirical)} method in that the conditional MLE of $\beta$ (which account for selection) is used to for estimating $\Sigma_{mc}$ (the method described in \S \ref{sect:sig_after_selection}). Additionally, we compare the correction methods to the \textit{Naive} and \textit{Full} methods. Adjustment for selection is done using PSAT (Algorithm \ref{alg:test}).

    Selection according to Tag-SNP is the same for all methods, as they rely on $X_o'y$ (see SM \S \ref{sect:equal}). The comparison highlights the gap in post selection inference when $X_o$ is unavailable. 
		
\begin{figure}[hbt!]
	 	\centering
	 	\includegraphics[width=0.95\textwidth]{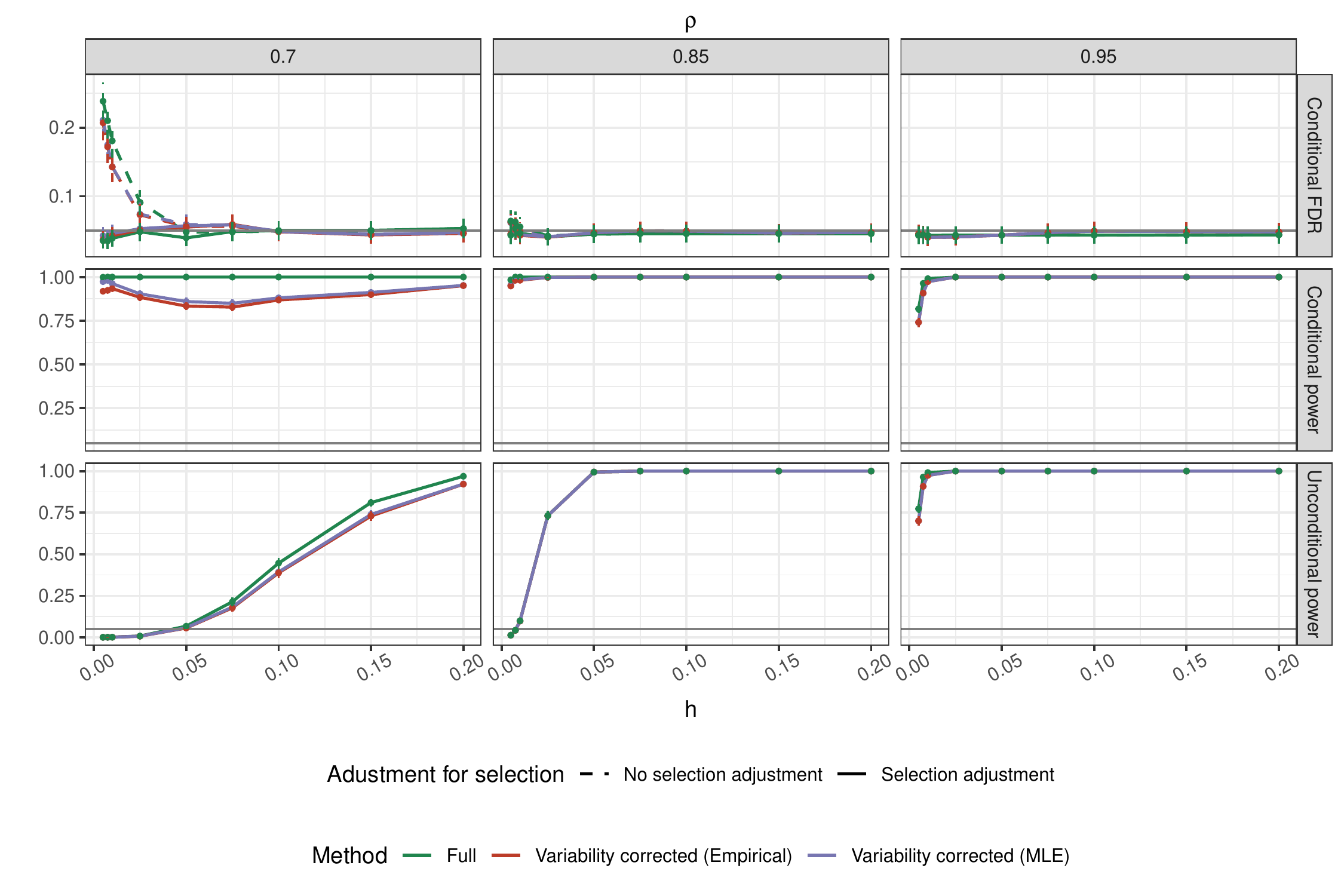}
	 	\caption{The conditional FDR, conditional power, and unconditional power as a function of $h$ (Eq. \eqref{eq:h}),  in settings mimicking genomic covariates, with variability correction and PSAT. The rows indicate the measure, the columns the correlation parameter $\rho$. $s^* = \{1,20\}$ and the Tag-SNP index is 10. The number of observation in the reference panel, $n_r = 1000$. The horizontal grey line is at the 0.05 nominal FDR level and the vertical lines represent the 2 standard error around the estimate.}
	 	\label{Fig:fine_psat}
	 \end{figure}

	We focus on the scenario where $s^* = \{1,20\}$, $n_r = 1000$, $h \in \{0.005, 0.01, 0.025, 0.05, 0.075, 0.1\}$ and $\rho \in \{0.75, 0.85, 0.95 \}$. In Figure \ref{Fig:fine_psat} (top row) the conditional FDR in the simulation configurations for all methods are presented. The difference seen between $\rho = 0.7$ and $\rho = 0.95$, is due to the selection process. In order to be selected, the Tag-SNP estimated coefficient must be above a certain threshold. As the correlation increases, this event becomes more likely (since the Tag-SNP is more correlated with SNPs in $s^*$). When the selection event happens with probability 1, no adjustment for selection is required, and the selection corrected methods and the non-adjusted selection method curves merge. Overall, the PSAT corrected procedures maintain the FDR.  
	There is almost no difference in terms of FDR between the \textit{Variability corrected (MLE)} and \textit{Variability corrected (Empirical)} methods. 
	
	Tag-SNP estimated marginal coefficient is above the threshold
	We consider conditional and unconditional power. The conditional power is the power to detect SNPs given that a selection event happened (only relying on iterations where the Tag-SNP estimated marginal coefficient is above the threshold).  For the unconditional power, we consider the number of times the region was selected and the causal SNPs were detected (relying on all of the iterations).
	
	The unconditional power decrease as $\rho$ increases. The increase in power is due to the selection event being more likely. As the correlation between the Tag-SNP and casual SNPs increase the selection event becomes more likely, increasing the unconditional power. This example demonstrate why the choice of the Tag-SNP is critical to obtain good power. 
	
    In the conditional power plot (Fig. \ref{Fig:fine_psat}, middle row), we see an interesting phenomena when $\rho = 0.7$. The conditional power curve of the selection adjusted procedure dips before continuing to rise as $h$ increases. 
	The dip in power occurs because when $h$ is small, the selection event occurs almost entirely at random (as evident by the unconditional power plots, right column). Furthermore, since $\rho = 0.7$, the causal SNPs are barely correlated with the Tag-SNP. Therefore, when the region is selected by chance, the adjustment required for the causal SNPs is very lenient. When $h$ increases, the selection event is driven by the causal SNPs; therefore, when testing them individually, the adjustment for the SNPs driving the selection is stricter. 	Finally,  the \textit{Variance corrected (MLE)} is as, or more powerful than the \textit{Variance corrected (Empirical)} method,  which is why we recommend to use it.

	\subsection{Real Genomic covariates} \label{sect:realSim}

    \subsubsection{UK-biobank simulation} \label{sect:uksim}

    In the following simulation the genomic data is taken from the UKbiobank data \citep{UKBiobank}, while the phenotype is simulated in order to investigate the method properties. In order to mimic dependencies found in GWAS studies we took a single SNP found to have marginal significance on BMI (using the GIANT consortium data) on Chromsome-19. All SNPs found in the UKbiobank data in distance less than 100kb are included in the simulation. 
    
    SNPs with MAF less than 0.05 were removed, and we randomly selected pairs of SNPs to avoid correlations larger than 0.97. Overall, 13 SNPs were used in the simulation. The SNPs are relatively spread across the Chromosome, leading to weaker correlations than those considered in the synthetic data simulation. 
    
    The construction of the phenotype is similar to that in the synthetic genotype simulation (\S \ref{sect:fabsim}). We considered $n_r \in \{ 500, 1500, 5000\}$, $n_o = 100,000$, and four sets of causal SNPs ($\{1,13\}$, $\{1,4,9,13\}$, $\{1,3,5,7,10, 13\}$ and $\{1,2, \ldots , 13\}$). We consider 3 methods, \textit{Naive}, \textit{Variability corrected (Empirical)} and \textit{Full}. In this simulation the region is pre-specified (i.e., not selected from the data using marginal screening). 
    
    \begin{figure}[hbt!] 
		\centering
		\includegraphics[width = 0.7\textwidth]{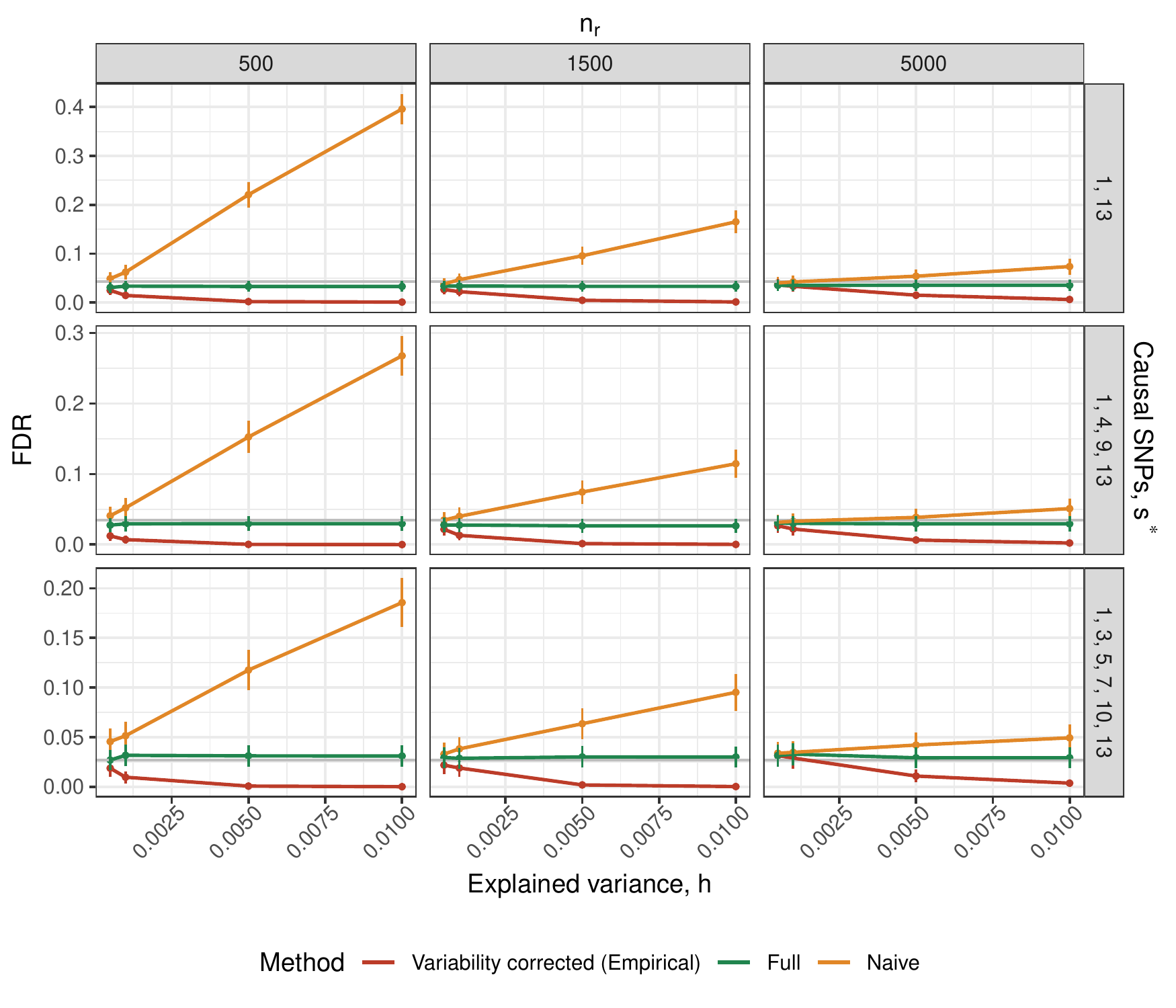}
		\caption{FDR results of UK-biobank based simulation. The rows are the set of causal SNPs, $s^*$, the columns are $n_r$. The x-axis is $h$ (Eq. \eqref{eq:h}) and $n_o = 100000$. Vertical lines represent the 2 standard error around the estimate.}
		\label{Fig:ukbiobank_sim_fdr}
	\end{figure}

    The results are consistent with the previous simulations finding. Figure \ref{Fig:ukbiobank_sim_fdr} demonstrates that the \textit{Naive} method fails in maintaining the expected FDR level. The FDR increases with the percentage of explained variance by the SNPs, and with the ratio of $\frac{n_o}{n_r}$. Furthermore, as the proportion of non-null SNPs increase so does the FDR. 

	\begin{figure}[hbt!] 
		\centering
		\includegraphics[width = 0.70\textwidth]{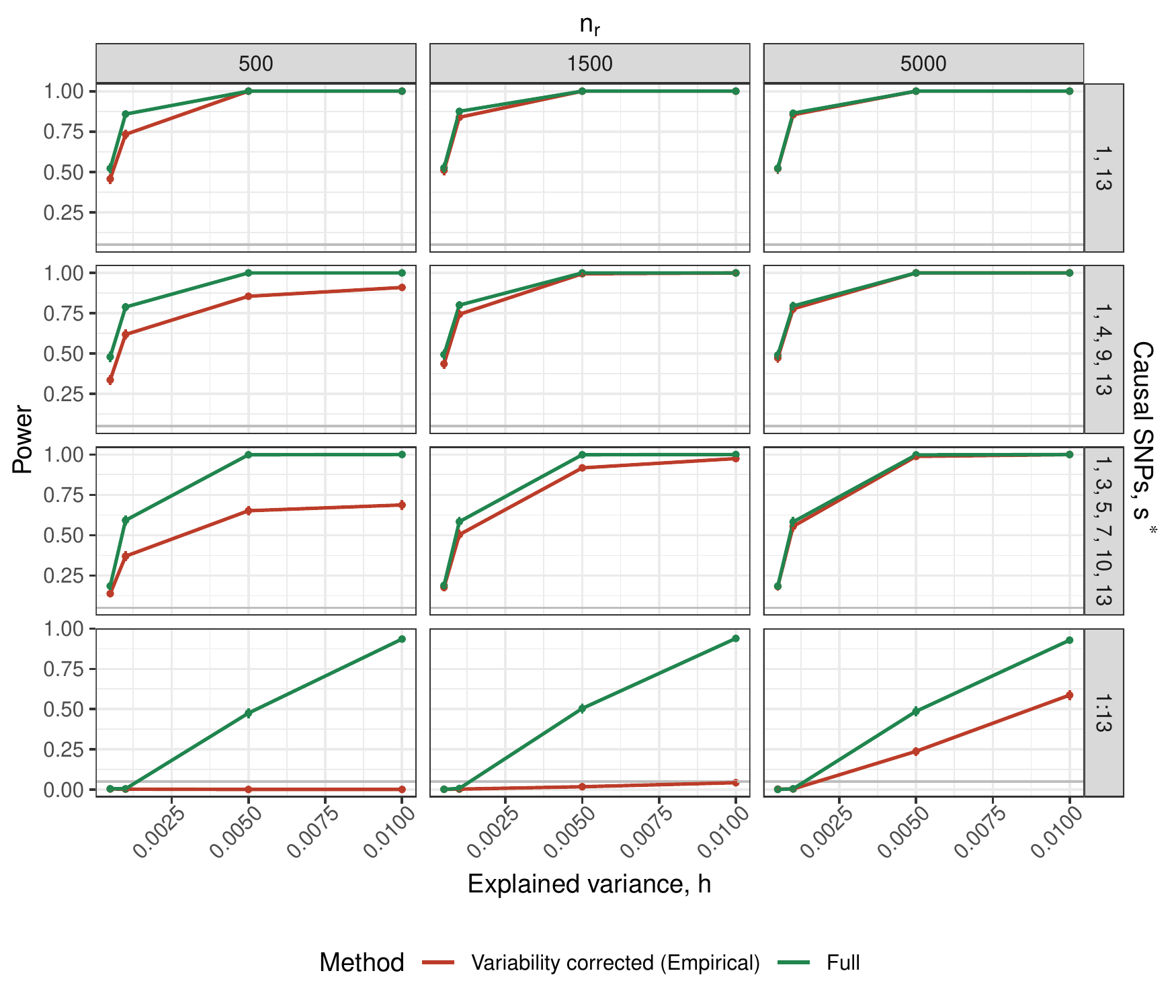}
		\caption{Power results of UK-biobank based simulation. The rows are the set of causal SNPs, $s^*$, the columns are $n_r$. The x-axis is $h$ (Eq. \eqref{eq:h}) and $n_o = 100000$. Vertical lines represent the 2 standard error around the estimate.}
		\label{Fig:ukbiobank_sim_power}
	\end{figure}

    In terms of power, the \textit{Variability corrected (Empirical)} method has low power when  $\frac{n_o}{n_r}$ is high compared to the coefficient size. In such cases the variance inflation is similar to the magnitude of the coefficient resulting in low power. This phenomena can be seen in Figure \ref{Fig:ukbiobank_sim_power} as the non-null proportion increase the power decrease, specifically for $n_r \in \{500, 1500\}$, when the non-null proportion is $1$ ($s^* = \{1,2,\ldots,13\}$) the power reduces to the specified FDR level, increasing only when $n_r$ increases to $5,000$.

	\subsubsection{1000Genome project data}
	
	The following simulation setting is similar to \S \ref{sect:fabsim}. A gene is analyzed only if the Tag-SNP estimated coefficient is above a certain threshold. However, the genotype is not simulated but rather taken from 1000Genome project.  We compare the variance estimation methods (coupled with PSAT) performance. 
    The gene analyzed is Lecithin–Cholesterol Acyltransferase (LCAT, rs:16:45409048 - 16:45412650 according to GRCh37). The SNPs were filtered to remove correlations higher than $0.99$ and MAF less than $0.05$, the genotype is taken from 1000Genome project. After filtration 5 SNPs are left, see Table \ref{tab:desc} for the correlation matrix and MAF. 
	
    \begin{table}[] \label{tab:desc}
    \centering
    	\begin{tabular}{cccccc}
    		\hline
    		Correlation & rs5923 & rs13336998 & rs13306496 & rs1109166 & rs11860115 \\
    		\hline
    		rs5923      & 1.00   & 0.69       & -0.01      & 0.48      & 0.79       \\
    		rs13336998  & 0.69   & 1.00       & 0.01       & 0.36      & 0.88       \\
    		rs13306496  & -0.01  & 0.01       & 1.00       & 0.40      & 0.02       \\
    		rs1109166   & 0.48   & 0.36       & 0.40       & 1.00      & 0.41       \\
    		rs11860115  & 0.79   & 0.88       & 0.02       & 0.41      & 1.00       \\
    		\hline 
    		MAF         & 0.15   & 0.07       & 0.11       & 0.66      & 0.09       \\ 
    		\hline
    	\end{tabular}
    	\caption{Correlation matrix of LCAT gene. Last row is the Minor Allele Frequency.}
    \end{table}

	To conduct the simulation, we split the sample of $n = 2504$ individuals so 80\%  are in the original study ($n_o = 2003$) and 20\% are in the reference panel ($n_r = 501$). The phenotype was generated artificially, with varying number of SNPs with non-zero coefficients, indexed as $\{ \{1\} ,\{1, 4\}, \{3, 5\} \}$, and explained variance $h \in \{0.05, 0.1, 0.15, 0.2, 0.3, 0.5, 0.75, 0.95\}$. We increase the range of $h$ to have greater power since we are limited in the number of observations. The Tag-SNP index is 3, and the threshold is $n_o \times Z_{1 - 0.05}$
		
	 \begin{figure}[!hbt]
	 	\centering
	 	\includegraphics[width=0.95\textwidth]{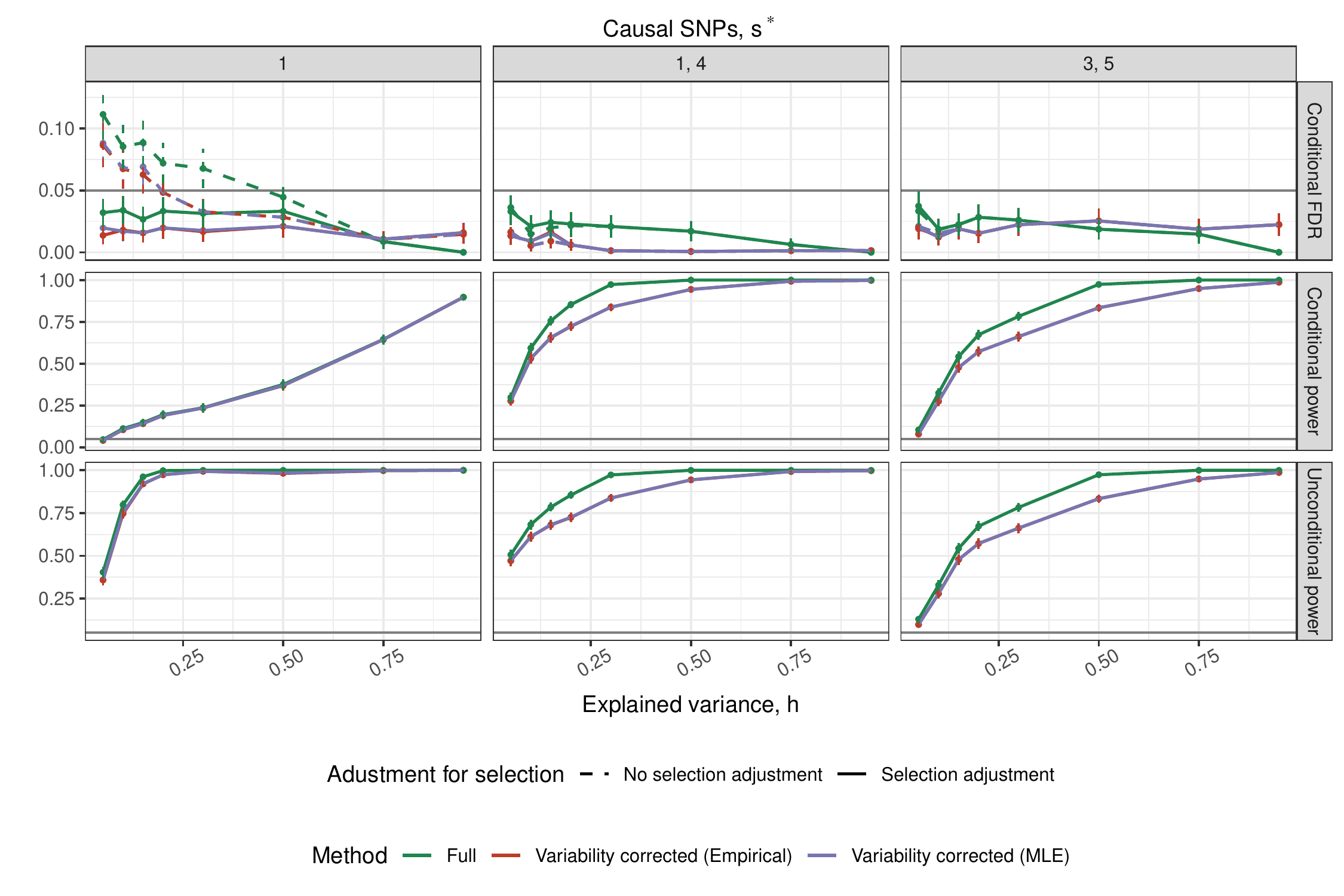}
 	\caption{Results on real genetic data and simulated phenotype, using variability correction with PSAT. The columns are the set of causal SNPs, $s^*$, the rows are the relevant measure. The x-axis is $h$ (Eq. \eqref{eq:h}), $n_r = 501$ and $n_o = 2003$. The grey line passes at the expected FDR level $0.05$. Vertical lines represent the 2 standard error around the estimate. }
	 	\label{Fig:real_fdr_psat}
	 \end{figure}
	 
	Figure \ref{Fig:real_fdr_psat} (top row) demonstrates that correcting for selection and the reference panel covariance estimation are required to maintain the conditional FDR. In the power analysis, methods that have not maintained the conditional FDR are dropped, leaving us with the \textit{Full}, \textit{Variance corrected (Empirical)} and \textit{Variance corrected (MLE)} methods. 
	 
	Both the conditional and unconditional power are inspected in Figure \ref{Fig:real_fdr_psat}. The PSAT correction becomes stricter the less likely the selection event is. When the Tag-SNP is one of the causal SNPs, $s^* = \{3,5\}$ or the casual SNP is correlated with the Tag-SNP $s^* = \{1, 4\}$, the selection happens with probability 1. So the conditional and unconditional power curves coincide, see Figure \ref{Fig:real_fdr_psat} (bottom and middle rows). 
	 
	In Figure \ref{Fig:real_fdr_psat} (bottom and middle rows), the power of the suggested method is on par with the \textit{Full} method. As the number of causal SNPs increases (so each SNP  explains less of the overall variability in the phenotype), the method's power decreases. The Tag-SNP's choice plays a crucial role in this example; when the Tag-SNP is not one of the causal SNPs (or not correlated to them, see $s^* = {1}$), the power decreases substantially for all methods. There seem to be no difference in power between the \textit{Variability corrected (Empirical)} and the \textit{Variability corrected (MLE)}.

    \section{Real data examples} 

    \subsection{Analysis of UK-biobank FTO gene} 
    
    To assess the usefulness of the suggested method on actual genomic and phenotype data, we used the UK-biobank data on the FTO gene located in Chromosome-16. It was found to be correlated with obesity according to several studies \citep{gerken2007obesity, frayling2007common}. We used the GIANT consortium data to find SNPS with marginal association of less than $5 \times 10^{-8}$ with BMI and included all SNPs in distance less than 100kb from them.  In pairs of SNPs with absolute correlation higher than 0.97, one was randomly chosen. Overall, 33 SNPs from the FTO gene were included in the simulation. 
    
    To assess validity, we repeatedly sampled $n_r \in \{500, 1000, 5000 \}$ and $n_o = 50000$ observations to serve as our reference and original study. The results are reported compared to the complete data containing 380,776 observations. 
    Three methods were compared, the \textit{Naive}, \textit{Variance corrected (Empirical)} and the \textit{Full}. The analysis was conducted at $\alpha = 0.05$, and p-values were adjusted using BH. Using the complete data, 20 SNPs were rejected, suggesting that the actual FDP should be less than $0.05$. The explained variance using the complete data-set is $0.0054$. All results are compared to the \textit{Complete} method. 
    
    The results are given in Table \ref{tab:FTO}. The suspected false discoveries of each method are the SNPs found significant by the method in question but not by the \textit{Complete} method. The average proportion of suspected false discoveries is the highest for \textit{Naive} method. The average proportion of false discoveries decrease as $n_r$ increase,  but still, for $n_r = 5000$ this proportion is higher than in the \textit{Full} method, serving as our baseline. 
    The \textit{Variability corrected} method is conservative, discovering on average only 8.6\% of SNPs discovered by the \textit{Complete} method. The proportion of discovered SNPs by both methods increases considerably when $n_r$ increases. When the proportion of non-null coefficients is large, the power of the \textit{Variability corrected} decreases. Further investigation to the relationship between power and proportion of non-null coefficients is explored in \S \ref{sect:uksim}. If the gene had explained a higher proportion of the BMI variance, it is expected that the \textit{Naive} method would have made even more suspicious false discoveries. In contrast, the \textit{Variability corrected} would have discovered more SNPs found by the \textit{Complete} method while maintaining a similar number of suspicious discoveries. 

    \begin{table}[]\label{tab:FTO}
    \centering
    \resizebox{\textwidth}{!}{
    \begin{tabular}{ccccccc} \hline 
     &  
      \multicolumn{3}{c}{\% (Not rejected by \textit{Complete} $\cap$ rejected by \textit{method})} & 
      \multicolumn{3}{c}{\% (Rejected by \textit{Complete} $\cap$ Rejected by \textit{method})} 
      \\  \hline 
      $n_r$ & 
      Naive &
      Variability corrected&
      Full &
      Naive &
      Variability corrected&
      Full 
       \\ \hline 
500  & 0.169 & 0.013 & 0.036 & 0.598 & 0.086 & 0.512 \\
1500 & 0.085 & 0.014 & 0.033 & 0.540 & 0.206 & 0.513 \\
5000 & 0.052 & 0.024 & 0.035 & 0.517 & 0.366 & 0.509 \\ \hline 
    \end{tabular}}
    \caption{Comparison of methods on FTO gene analysis containing 33 SNPs. The \textit{Naive} method reject a higher percentage of SNPs not rejected by the complete method (using all of the observations). Note, the parameter $n_r$ has no effect on the \textit{Full} method, and the differences between rows are only due to different samples taken from the original study.}
    \end{table}

	\subsection{Application to variant selection following gene-level testing}
	
	We apply the suggested method on the data of the Dallas Heart Study (DHS) \citep{romeo2007population}. The data consists of 3549 individuals (601 Hispanic, 1830 non-Hispanic black, 1043 non-Hispanic white, and 75 other ethnicities), and four genes were considered. 
	
    The phenotype in question is Triglyceride (TG). The response variable is the residuals of log(TG) after adjusting for race, sex, and age. The data was split to original study $n_o = 3000$ and the rest of the observations were used as the reference panel, $n_r = 549$. The data split was repeated until each variant appeared at least once in the reference and original study, excluding variants that have a mutation only in a single subject. 

    The quadratic SKAT test \citep{wu2011rare} is used as the global-null test to select the relevant genes. As in the study of \cite{romeo2007population}, four genes are tested, and only ANTPGL4 is selected if the SKAT (p-value $< 0.05 / 4$). After filtering the variants, we remain with 11 out of the original 32. 
    The rest of the analysis is as described in Algorithm \ref{alg:test}. It is compared to the PSAT analysis conducted using the original data.  
    
    We present the results in Table \ref{tab:DHS}. The estimated coefficients of the analysis using the summary level data and reference panel are compared with the original study analysis. The results are similar with both the coefficients and p-values. After applying BH correction, both methods find E40K, R278Q, and V308M as significant variants.

\begin{table}[] \label{tab:DHS}
\resizebox{\textwidth}{!}{
\begin{tabular}{ccccccccc}
    \hline
    \multirow{2}{*}{} & \multicolumn{2}{c}{\# of rare variants} & \multicolumn{2}{c}{Summary level data analysis} & \multicolumn{2}{c}{Original study analysis} &  &  \\
    \hline
             Variant             & Original study  & Reference panel  & Estimated Coefficients               & P-value               & Estimated Coefficients             & P-value             &  &  \\
                          \hline
    P5L                   & 1               & 1                & 0.00919                 & 0.65571               & 0.00814               & 0.61269             &  &  \\
    E40K                  & 41              & 9                & -0.04951                & 0.00381               & -0.05371              & 0.00699             &  &  \\
    M41I                  & 21              & 7                & 0.02572                 & 0.17311               & 0.02489               & 0.15691             &  &  \\
    S67R                  & 1               & 1                & 0.01478                 & 0.45246               & 0.01372               & 0.41595             &  &  \\
    R72L                  & 2               & 1                & -0.00735                & 0.64675               & -0.00837              & 0.68566             &  &  \\
    E190Q                 & 24              & 8                & 0.02104                 & 0.34591               & 0.01732               & 0.24879             &  &  \\
    T266M                 & 1581            & 306              & 0.01687                 & 0.24034               & 0.02280               & 0.36533             &  &  \\
    R278Q                 & 185             & 22               & -0.05618                & 0.00308               & -0.05423              & 0.00216             &  &  \\
    V308M                 & 2               & 1                & 0.06126                 & 0.00075               & 0.06159               & 0.00074             &  &  \\
    R336C                 & 6               & 1                & 0.00292                 & 0.85407               & 0.00336               & 0.87213             &  &  \\
    G361R                 & 1               & 1                & 0.00933                 & 0.65040               & 0.00828               & 0.60749             &  & \\ \hline
    \end{tabular}
    }
    \caption{Results of analysis on 11 variants (with more than one carrier) of ANGPTL4. Column 1 and 2 show the number of variants in each sample, columns 3 and 4 the estimated coefficients according to original split and reference split and column 5,6 are the p-values. For both methods the selection was conducted using SKAT test at significance level of 0.05/4. In both methods the variants found signifcant at level 0.05 after BH correction are E40K, R278Q and V308M.}
    \end{table}

	\section{Discussion}

    
    We show that using $X_r$ instead of $X_o$ requires adjustment for valid inference and suggest a method for adjusting for it. 
    A major setback of the suggested method is its computational complexity. We suggest two ways to alleviate the burden (1) if $X^*_{i,.}$ is approximately multivariate Gaussian, then the complexity is significantly reduced. (2) use a threshold estimate for $\beta$, so if many entries are zero, computing $\hat{\boldsymbol{V}}_{\hat{R}} $ has lower complexity. Furthermore, improvement to the computational complexity can come from a more efficient estimation of the sample covariance matrix variance. The variance estimation procedure complexity is $O(n p^4)$, which is the method's main computational bottleneck. The estimation, however, can be easily parallelized, reducing the computation time. 
	
    The method has been developed for a continuous phenotype. Generalizing it to the binary phenotype is not straightforward, since estimating the covariance between the coefficients of logistic regression relies on the estimation of the phenotype probability for each individual in the reference panel -  information that we do not have access to from marginal regression summary statistics. 
    
    A different interesting application of the method is meta-analysis, where more than one study is involved. In such case, the application of the method is more straightforward (as long as the phenotype of interest is continuous). Furthermore, the problem of sharing genotype level data in meta-analysis is exacerbated as more individuals are involved from a larger number of centers.
    
    Another extension is the use of  simulated reference panels. Recent applications of knockoffs \cite{barber2015controlling}   requires the estimation of the joint distribution of the covariates. If the  joint distribution is approximately known,  the reference sample can be increased artificially by sampling.
    If the reference panel is of similar size to the original study (and both are large) then empirically the variance inflation does not affect the FDR by much (see Fig. \ref{Fig:fine_fdr_gauss} 1, $n_r = 10000$). Thus, if the reference sample can be artificially increase to similar size to that of the original study sample size the \textit{Naive} method can be relatively safe to use.

\bibliographystyle{chicago}
\bibliography{refrence}


\section*{Supplementary material}

# Supplementary Material: Inferring on joint associations from marginal associations and a reference sample

## S1   Proof of Theorem 3.1

We aim to estimate the distribution of $\hat{\beta}_{mc} = \frac{n_r}{n_o}(X'_r X_r)^{-1} X'_o y_o$. Where $X_r, X_o$ and $y_o$ are normalized. Assume that $X^*_{1,.}, X^*_{2,.}, \ldots, X^*_{n_o+n_r,.}$ are sampled from a distribution with expected value, $\mathbb{E}(X_i) = \mu$, and variance $\mathbb{V}(X_i) = \Sigma$, and it has a bounded fourth moment. So with out loss of generality that the first $n_o$ observation consist of $X^*_o$ and the last $n_r$ observation form $X^*_r$. Let $\hat{\Sigma}_k = n_k^{-1} \sum_{i=1}^{n_k} (X_{k,i} - \bar{X}_k)(X_{k,i} - \bar{X}_k)'$. Define $X_{k,i,.} = d(\hat{\Sigma}_k)^{-0.5}(X^*_{k,i,.} - \bar{X}^*_k)$, $k = \{o, r\}$.

From the law of large numbers as $n_k \to \infty$, then $\bar{X}^* \xrightarrow{a.s.} \mu$, and $\hat{\Sigma}_k \xrightarrow{a.s.} \Sigma$, $k \in \{o, r\}$. According to Slutskys' theorem

$$X_{i,.} \xrightarrow{d} d(\Sigma)^{-0.5}(X^*_{i,.} - \mu).$$

Before we turn to prove Theorem 3.1, we provide some necessary background.

Since the proof involve estimating the variance of sample covariance matrices we utilize matrix operators and transformations, which are briefly described here. We follow the same notations as in Neudecker and Wesselman (1990): $vec$ is the vectorization operator: $vec(K)$ concatenates the columns of matrix $A \in \mathbb{R}^{m \times n}$ to a vector $vec(A) \in \mathbb{R}^{mn}$. Let $C \in \mathbb{R}^{p \times q}$. The Kronecker product, $\otimes$, is defined as follows $A \otimes C$ is a matrix of size $mp \times nq$ of the following form,

$$A \otimes C = [a_{i,j}C] = \begin{bmatrix} a_{1,1}C & \ldots & a_{1,n}C \\ \vdots & \ddots & \vdots \\ a_{m,1}C & \ldots & a_{m,n}C \end{bmatrix},$$

where $a_{i,j}$ is the $(i,j)$ entry of $A$.

A useful result that combines $vec$ and $\otimes$ is

$$vec(ABC) = (C' \otimes A)vec(B) = (C'B' \otimes I_m)vec(A), \tag{1}$$

where $B \in \mathbb{R}^{n \times p}$. Another useful property of the Kroneker product is the mixed product property for matrices $A, C$ and $B, D$ of suitable dimensions,

$$(A \otimes B)(C \otimes D) = (AC) \otimes (BD). \tag{2}$$

The variance of $\sqrt{n_r}\hat{R}_r$, $\boldsymbol{V}_{\hat{R}}$, is given in Theorem 2 of Neudecker and Wesselman (1990), reproduced here, due to its importance for our theorem.

**Theorem S1.1.** *[Theroem 2 - Neudecker and Wesselman (1990)] Let $\{X_{i,.}^* : i = 1, \ldots n\}$ be a set of $n$ i.i.d random vectors with correlation matrix $R$, covariance matrix $\Sigma$ and sample covariance and correlation matrices $\hat{\Sigma}$ and $\hat{R}$. Then,*

$$\lim_{n \to \infty} \sqrt{n} \times vec(\hat{R} - R) \stackrel{d}{=} N_{p^2}\left(0, \boldsymbol{V}_{\hat{R}}\right),$$

*where*

$$\boldsymbol{V}_{\hat{R}} = \left(I - M_s(I \otimes R)M_d\right)\left(d(\Sigma)^{-0.5} \otimes d(\Sigma)^{-0.5}\right)\boldsymbol{V}_{\hat{\Sigma}} \\ \left(d(\Sigma)^{-0.5} \otimes d(\Sigma)^{-0.5}\right)\left(I - M_d(I \otimes R)M_s\right). \tag{3}$$

$\mathbf{V}_{\hat{\Sigma}}$ *is defined as*

$$\mathbf{V}_{\hat{\Sigma}} = \mathbb{E}\left((X_{i,.} - \mu)(X_{i,.} - \mu)' \otimes (X_{i,.} - \mu)(X_{i,.} - \mu)'\right) - vec(\Sigma)vec(\Sigma)',$$

*and $M_s = 0.5(I + U)$, $U$ being the permutation matrix and $M_d = \sum_{i=1}^{p}(E_{i,i} \otimes E_{i,i})$. $E_{i,i}$ is a $p \times p$ matrix with 1 in the iith position and zero elsewhere.*

Using the Theorem, we can derive the asymptotic distribution of the inverse correlation matrix, given in the following lemma.

**Lemma 1.** *Under the same conditions of Theorem S1.1. Then,*

$$\lim_{n \to \infty} \sqrt{n} \times vec(\hat{R}^{-1} - R^{-1}) \sim N_{p^2}\left(0, (R^{-1} \otimes R^{-1})\boldsymbol{V}_{\hat{R}}(R^{-1} \otimes R^{-1})\right). \tag{4}$$

The proof is given in the §S1.2.

We can now turn to prove Theorem 3.1, we begin by finding the expectation and variance of

$$\hat{\beta}_{mc} = n_o^{-1}\hat{R}_r^{-1}X_o'(X_o\beta + \epsilon) = \hat{R}_r^{-1}\hat{R}_o\beta + n_o^{-1}\hat{R}_r^{-1}X_o'\epsilon. \tag{5}$$

The variance is,

$$
\begin{aligned}
\mathbb{V}\left(\hat{\beta}_{mc}\right) &= \mathbb{V}\left(\hat{R}_r^{-1}\hat{R}_o\beta + \frac{1}{n_o}\hat{R}_r^{-1}X_o\epsilon\right) \\
&= \mathbb{E}\left(\mathbb{V}\left(\hat{R}_r^{-1}\hat{R}_o\beta + \frac{1}{n_o}\hat{R}_r^{-1}X_o'\epsilon|X_o,X_r\right)\right) \\
&\quad + \mathbb{V}\left(\mathbb{E}\left(\hat{R}_r^{-1}\hat{R}_o\beta + \frac{1}{n_o}\hat{R}_r^{-1}X_o'\epsilon|X_o,X_r\right)\right) \\
&= \mathbb{E}\left(\frac{1}{n_o^2}\hat{R}_r^{-1}X_o'X_o\hat{R}_r^{-1}\sigma^2\right) + \mathbb{V}\left(\hat{R}_r^{-1}\hat{R}_o\beta\right) \\
&= \frac{\sigma^2}{n_o}R^{-1} + \mathbb{V}\left(\hat{R}_r^{-1}\hat{R}_o\beta\right) + O\left(\frac{1}{n_o n_r}\right).
\end{aligned}
\tag{6}
$$

The second equality follow from the law of total variance, and the last equality follows from Lemma 2.

**Lemma 2.** *Following the assumptions of Thereom 3.1.*

$$\mathbb{E}\left(\frac{1}{n_o^2}\hat{R}_r^{-1}X_o'X_o\hat{R}_r^{-1}\sigma^2\right) = \frac{\sigma^2}{n_o}R^{-1} + O\left(\frac{1}{n_o n_r}\right)$$

See proof in Section S1.1.

To simplify $\mathbb{V}\left(\hat{R}_r^{-1}\hat{R}_o\beta\right)$, we decompose $\hat{R}_r^{-1}\hat{R}_o\beta$ as follows:

$$
\begin{aligned}
\hat{R}_r^{-1}\hat{R}_o\beta &= \left(\hat{R}_r^{-1} - R^{-1} + R^{-1}\right)\left(\hat{R}_o - R + R\right)\beta = \\
&\left(\hat{R}_r^{-1} - R^{-1}\right)\left(\hat{R}_o - R\right)\beta + \left(\hat{R}_r^{-1} - R^{-1}\right)R\beta + R^{-1}\left(\hat{R}_o - R\right)\beta + \beta.
\end{aligned}
\tag{7}
$$

Since $\hat{R}_r$ and $\hat{R}_o$ are consistent estimators of $R$, and $R$ is invertible, $\left(\hat{R}_r^{-1} - R^{-1}\right)\left(\hat{R}_o - R\right)\beta$ is a magnitude smaller than $\left(\hat{R}_r^{-1} - R^{-1}\right)R\beta$ and $R^{-1}\left(\hat{R}_o - R\right)\beta$. Note, the covariance terms involving $\left(\hat{R}_r^{-1} - R^{-1}\right)\left(\hat{R}_o - R\right)$, are also $O\left(\frac{1}{n_o n_r}\right)$. For ease of notation we define $D_r = \hat{R}_r^{-1} - R^{-1}$ and $D_o = (\hat{R}_o - R)$. We show this for $Cov(D_r D_o \beta, R^{-1} D_o \beta)$:

$$
\begin{aligned}
Cov(D_r D_o \beta, R^{-1} D_o \beta) &= \mathbb{E}(D_r D_o \beta \beta' R^{-1} D_o) - \mathbb{E}(D_r D_o \beta)\mathbb{E}(R^{-1} M_o \beta)' \\
&= \mathbb{E}(D_r)\mathbb{E}(D_o \beta \beta' D_o) - \mathbb{E}(D_r)E(D_o)\beta \beta' \mathbb{E}(D_o)R^{-1} \\
&= O\left(\frac{1}{n_o n_r}\right).
\end{aligned}
\tag{8}
$$

The second equality is due to independence of $D_r$ and $D_o$, the third is since $\mathbb{E}(D_o \beta) = O\left(\frac{1}{n_o}\right)$ and $\mathbb{E}(D_r) = O\left(\frac{1}{n_r}\right)$, and $\mathbb{E}(D_o \beta \beta' D_o) = \mathbb{V}(D_o \beta) = O\left(\frac{1}{n_o}\right)$.

Similar arguments can be employed to the other covariance term, $cov(M_r M_o \beta, M_r R^{-1})$ yielding the result. Due to $\hat{R}_r$ and $\hat{R}_o$ being independent, we can write:

$$
\mathbb{V}(\hat{R}_r^{-1}\hat{R}_o \beta) = \mathbb{V}\left(\hat{R}_r^{-1} R \beta\right) + \mathbb{V}\left(R^{-1}\hat{R}_o \beta\right) + O\left(\frac{1}{n_o n_r}\right).
\tag{9}
$$

We shall approximate the two terms in the right hand side of Eq. (9), as follows. We begin with the variance of $\hat{R}_r^{-1}$, for which the asymptotic distribution is given in Lemma 1,

$$
\lim_{n_r \to \infty} \sqrt{n_r} \times vec(\hat{R}_r^{-1} - R^{-1}) \sim N_{p^2}\left(0, (R^{-1} \otimes R^{-1})\boldsymbol{V}_{\hat{R}}(R^{-1} \otimes R^{-1})\right).
\tag{10}
$$

Therefore,

$$\lim_{n_r \to \infty} n_r \mathbb{V}\left(\hat{R}_r^{-1} R \beta\right) = \lim_{n_r \to \infty} n_r \mathbb{V}\left[vec(\hat{R}_r^{-1} R \beta)\right]$$

$$= \lim_{n_r \to \infty} n_r \mathbb{V}\left[(\beta' R' \otimes I)vec(\hat{R}_r^{-1})\right]$$

$$= \lim_{n_r \to \infty} n_r (\beta' R \otimes I)\mathbb{V}(vec(\hat{R}_r^{-1}))(R\beta \otimes I)$$

$$= (\beta' R \otimes I)(R^{-1} \otimes R^{-1})\boldsymbol{V}_{\hat{R}}(R^{-1} \otimes R^{-1})(R\beta \otimes I)$$

$$= (\beta' \otimes R^{-1})\boldsymbol{V}_{\hat{R}}(\beta \otimes R^{-1}).$$

The second equality follows from Eq. (1), the fourth from Eq. (10) and the last from the mixed product property Eq. (2). We obtain,

$$\lim_{n_r \to \infty} \sqrt{n_r}(\hat{R}_r^{-1} R \beta - \beta) \to N_p\left(0, (\beta' \otimes R^{-1})\boldsymbol{V}_{\hat{R}}(\beta \otimes R^{-1})\right) \tag{11}$$

The second term, $R^{-1}\hat{R}_o \beta$ asymptotic distribution is,

$$\lim_{n_o \to \infty} \sqrt{n_o}(R^{-1}\hat{R}_o \beta - \beta) \to N_p\left(0, (\beta' \otimes R^{-1})\boldsymbol{V}_{\hat{R}}(\beta \otimes R^{-1})\right), \tag{12}$$

since the asymptotic variance is,

$$\lim_{n_o \to \infty} n_o \mathbb{V}\left(R^{-1}\hat{R}_o \beta\right) = \lim_{n_o \to \infty} n_o \mathbb{V}\left((\beta' \otimes R^{-1})vec(\hat{R}_o)\right)$$

$$= (\beta' \otimes R^{-1})\boldsymbol{V}_{\hat{R}}(\beta \otimes R^{-1}),$$

where the first and second equalities are due to Eq. (1) and Eq. (12) respectively.

Now we can find the asymptotic variance of $\hat{\beta}_{mc}$

$$\lim_{n_r,n_o \to \infty} n_r \mathbb{V}(\hat{\beta}_{mc}) = \lim_{n_r,n_o \to \infty} n_r \left(\frac{\sigma^2}{n_o} R^{-1} + \mathbb{V}\left(\hat{R}_r^{-1} R \beta\right) + \mathbb{V}\left(R^{-1}\hat{R}_o \beta\right) + O\left(\frac{1}{n_r n_o}\right)\right)$$

$$= \lim_{n_r,n_o \to \infty} \left(\frac{\sigma^2 n_r}{n_o} R^{-1} + \frac{n_r + n_o}{n_r n_o}(\beta' \otimes R^{-1})\boldsymbol{V}_{\hat{R}}(\beta \otimes R^{-1})\right),$$

$$\tag{13}$$

The first equality is due to Eq. (6) and (9), and second equality is due to Eq. (11) and Eq. (12). Assuming that $\lim_{n_r, n_o \to \infty} \frac{n_r}{n_o} = c, \ c \in (0,1)$, then the asymptotic variance is,

$$\lim_{n_r, n_o \to \infty} n_r \mathbb{V}(\hat{\beta}_{mc}) = c\sigma^2 R^{-1} + (1+c)(\beta' \otimes R^{-1}) \boldsymbol{V}_{\hat{R}}. \tag{14}$$

Before giving $\hat{\beta}_{mc}$ we are left to show it is consistent,

$$\begin{aligned}
\lim_{n_r, n_o \to \infty} (\hat{\beta}_{mc}) &= \lim_{n_r, n_o \to \infty} \hat{R}_r^{-1} \hat{R}_o \beta + \lim_{n_r, n_o \to \infty} n_o^{-1} \hat{R}_r^{-1} X_o' \epsilon \\
&= \lim_{n_r, n_o \to \infty} \hat{R}_r^{-1} \hat{R}_o \beta = \beta,
\end{aligned} \tag{15}$$

where the second equality is due to $\mathbb{E}\left(n_o^{-1} \hat{R}_r^{-1} X_o' \epsilon\right) = 0$, and the third is due to Eq. 7.

Since each element of the decomposed $\hat{\beta}_{mc}$ is asymptotically distributed as normal (see Eq. (11) and Eq. (12)) and according to Eq. (14) then,

$$\lim_{n_r, n_o \to \infty} \sqrt{n_r}(\hat{\beta}_{mc} - \beta) \sim N_p(0, c\sigma^2 R^{-1} + (1+c)(\beta' \otimes R^{-1}) \boldsymbol{V}_{\hat{R}}(\beta \otimes R^{-1})).$$

## S1.1   Proof of Lemma 2

The expected value of $E\left(\frac{1}{n_o^2} \hat{R}_r^{-1} X_o' X_o \hat{R}_r^{-1} \sigma^2\right)$

$$\begin{aligned}
\mathbb{E}\left(e_i' \hat{R}_r^{-1} \hat{R}_o \hat{R}_r^{-1} e_j\right) &= \mathbb{E}\left(\mathbb{E}\left(e_i' \hat{R}_r^{-1} \hat{R}_o \hat{R}_r^{-1} e_j | X_r\right)\right) \\
&= \mathbb{E}\left(e_i' \hat{R}_r^{-1} R \hat{R}_r^{-1} e_j\right) \\
&= \mathbb{E}\left(tr\left(e_i' \hat{R}_r^{-1} R \hat{R}_r^{-1} e_j\right)\right) \\
&= tr\left(\mathbb{E}\left(\hat{R}_r^{-1} e_j e_i' \hat{R}_r^{-1}\right) R\right) \\
&= tr\left(\left(Cov\left(\hat{R}_r^{-1} e_j, e_i' \hat{R}_r^{-1}\right) + R^{-1} e_j e_i' R^{-1}\right) R\right) \\
&= tr\left(Cov\left(\hat{R}_r^{-1} e_j, e_i' \hat{R}_r^{-1}\right) R\right) + \left(R^{-1}\right)_{i,j}.
\end{aligned} \tag{16}$$

Where $e_i$ is a zero vector except for position $i$ where the entry is 1. The fifth equality follow due to $\mathbb{E}(R^{-1}e_j e_i' R^{-1}) = cov(R^{-1}e_j, e_i' R^{-1}) + \mathbb{E}(R^{-1}e_j)\mathbb{E}(e_i' R^{-1'})$. Since the asymptotic variance of $\hat{R}_r^{-1}$ tends to 0 at a rate of $n_r^{-1}$ (see Eq. (10)), $Cov\left(\hat{R}_r^{-1}e_j, e_i' \hat{R}_r^{-1}\right)$ is $O(n_r^{-1})$ assuming a bounded fourth moment of $X_i^*$, multiplying by $\frac{\sigma^2}{n_o}$ yields the result.

## S1.2   Proof of Lemma 1

According to Chapter 18, Eq. (12) of Magnus and Neudecker (2019),

$$dR^{-1} = -R^{-1}(\mathrm{d}R)R^{-1},$$

where d is a differential, a small quantity often used in matrix calculus to simplify derivative calculations (see same chapter for further details).

Applying $vec$ to both sides of the equation,

$$vec(\mathrm{d}R^{-1}) = -vec(R^{-1}(\mathrm{d}R)R^{-1}) = -(R^{-1} \otimes R^{-1})vec(\mathrm{d}R),$$

the last equality follows from Eq. (1). According to Theorem 18.1 (tying differentials and derivatives) in Magnus and Neudecker (2019) and linearity of the derivative,

$$\frac{\mathrm{d}vec(R^{-1})}{\mathrm{d}vec(R)} = \frac{\partial vec(R^{-1})}{\partial vec(R)'} = -(R^{-1} \otimes R^{-1}).$$

Applying the delta method, we obtain the result,

$$\lim_{n\to\infty} \sqrt{n}vec(\hat{R}^{-1} - R^{-1}) \sim N_{p^2}\left(0, (R^{-1} \otimes R^{-1})\boldsymbol{V}_{\hat{R}}(R^{-1} \otimes R^{-1})\right).$$

# S2   Computing an estimator of the variance

The expression in Eq. (14) needs to be estimated, to do so we replace all of the parameters with their estimators. However, the expression is still computationally complex to evaluate. In the following section we will describe the computation. We begin with Eq. 3, denote $P = \left(I - M_s(I \otimes R)M_d\right)$. We can rewrite Eq. 3 as $\boldsymbol{V}_R = P\mathbf{V}_{\hat{\Sigma}}P'$.

$$P = \left( I - M_s(I \otimes R)M_d \right) = \left( I - 0.5(I + K)(I \otimes R) \sum_{i=1}^{p}(E_{i,i} \otimes E_{i,i}) \right)$$

$$= I - 0.5 \left( \sum_{i=1}^{p}(E_{i,i} \otimes RE_{i,i}) - (RE_{i,i} \otimes E_{i,i}) \right).$$

Notice, that $RE_{i,i}$ is a matrix of all zeros except for the $i$'th column which is $R_{.,i}$. For each Kronecker multiplication of a two $p$ by $p$ matrices, we can look at an $i,j$'th block which is a $p \times p$ matrix, we denote this block by square brackets. That is the $A[i,j]$, $A \in \mathbb{R}^{p^2 \times p^2}$ denotes the submatrix of the elements in rows $i \times p$ to $(i + 1) \times p$ and columns $j \times p$ to $(j + 1) \times p$.

$$\sum_{i=1}^{p}(E_{l,l} \otimes RE_{l,l})[i,j] = \begin{cases} 0, & i \neq j \\ RE_{i,i}, & i = j \end{cases}$$

and

$$\sum_{l=1}^{p}(RE_{l,l} \otimes E_{l,l})[i,j] = E_{j,j}RE_{i,j}.$$

Thus, the matrix we are interested in can be written as,

$$P[i,j] = \begin{cases} I - 0.5RE_{i,i} - 0.5E_{i,i}RE_{i,i} & i = j \\ -0.5E_{j,j}RE_{i,j} & i \neq j \end{cases}.$$

The second term we are interested in is, $\Lambda = \left( (d(\Sigma))^{-0.5} \otimes (d(\Sigma))^{-0.5} \right)[i,j]$,

$$\Lambda[i,j] = \begin{cases} \sigma_{i,i}^{-0.5} \left( d(\Sigma) \right)^{-0.5} & i = j \\ 0 & i \neq j \end{cases}.$$

The multiplication of these two sparse sub-matrices is $O(p^2)$ (Park et al., 1992), as at most, the number of non-zero elements of $P[i,j]$ and $\left( (d(\Sigma))^{-0.5} \otimes (d(\Sigma))^{-0.5} \right)[i,j]$ is $2p$. The remaining matrix, $\mathbf{V}_{\hat{\Sigma}}$, can be partitioned as following

$$\mathbf{V}_{\hat{\Sigma}}[i,j] = Cov(\hat{\Sigma}_{i,.}, \hat{\Sigma}_{j,.}).$$

To deal with the expression $(\hat{\beta}' \otimes \hat{R}^{-1})$ in Eq. 14, we threshold $\hat{\beta}$ such

$$\hat{\beta}_i^* = \begin{cases} \hat{\beta}_{mc_i} & |\hat{\beta}_{mc_i}| \geq t \\ 0 & |\hat{\beta}_{mc_i}| < t \end{cases}.$$

Therefore, we rewrite the term as $(\hat{\beta}^{*'} \otimes \hat{R}_r^{-1})\hat{\mathbf{V}}_R(\hat{\beta}^{*'} \otimes \hat{R}_r^{-1})$.

$$(\hat{\beta}^{*'} \otimes \hat{R}_r^{-1})[1,j], = \begin{cases} \hat{\beta}_{mc_j}\hat{R}_r^{-1} & |\hat{\beta}_{mc_j}| \geq t \\ 0 & |\hat{\beta}_{mc_j}| < t \end{cases}.$$

$t$ is some prespecified threshold, the Kronecker multiplication is denoted by a single index, as there are only $p$ sub-matrices. We turn to write the expression of Eq. (14), as block matrix multiplication. Since all blocks are $p \times p$, the matrices are conformable,

$$(\hat{\beta}^{*'} \otimes \hat{R}_r^{-1})\hat{\boldsymbol{V}}_R(\hat{\beta}^{*'} \otimes \hat{R}_r^{-1}) = \sum_{i=1}^p \sum_{j=1}^p (\hat{\beta}^{*'} \otimes \hat{R}_r^{-1})[1,i]\hat{\boldsymbol{V}}_{\hat{R}}[i,j](\hat{\beta}^* \otimes \hat{R}_r^{-1})[j,1].$$

We assume w.l.o.g, that $\hat{\beta}^*$ has $m$ non-zero elements that are positioned at the beginning of the vector (i.e in locations $1, \ldots, m$ of the vector $\hat{\beta}^{*'}$),

$$\sum_{i=1}^p \sum_{j=1}^p (\hat{\beta}^{*'} \otimes \hat{R}_r^{-1})[1,i]\hat{\boldsymbol{V}}_{\hat{R}}[i,j](\hat{\beta}^* \otimes \hat{R}_r^{-1})[j,1] = \hat{R}_r^{-1} \left( \sum_{i=1}^m \sum_{j=1}^m \hat{\beta}_{mc_i}\hat{\beta}_{mc_j}\hat{\boldsymbol{V}}_{\hat{R}}[i,j] \right) \hat{R}_r^{-1}.$$

Thresholding reduces the amount of operation from $p^2$ to $m^2$. However, we still need to find $\hat{\boldsymbol{V}}_R[i,j]$, again writing as block matrix multiplication we obtain,

$$\hat{\boldsymbol{V}}_R[i,j] = P\Lambda \hat{\mathbf{V}}_{\hat{\Sigma}}\Lambda P'$$
$$= \sum_{l=1}^{p}\sum_{k=1}^{p}(P\Lambda)[i,l]\hat{\mathbf{V}}_{\hat{\Sigma}}[l,k](\Lambda P')[k,j].$$

The final computation is,

$$\hat{R}^{-1}\left(\sum_{i=1}^{m}\sum_{j=1}^{m}\hat{\beta}_{mc_i}\hat{\beta}_{mc_j}\sum_{l=1}^{p}\sum_{k=1}^{p}(P\Lambda)[i,l]\hat{\mathbf{V}}_{\hat{\Sigma}}[l,k](\Lambda P')[k,j]\right)\hat{R}^{-1}.$$

Estimating the covariance matrix of $vec\left(\hat{\Sigma}\right)$, is $o(np^4)$. The complexity of estimating the variance of $\hat{\beta}_{mc}$ is $o(m^2p^2 + np^4)$. Notice, that if $X$ is a assumed to be normal then $\mathbb{V}\left(vec\left(\hat{\Sigma}\right)\right) = 2M_s(\Sigma \otimes \Sigma)$, since we plug in $\hat{\Sigma}$ it reduces the computation time to $o(m^2p^2 + p^4)$.

# S3 Equivalence of testing marginal coefficients

In the following section we show the equivalence of testing the Tag-SNP marginal coefficient in both Full and Variability corrected methods. The PSAT framework requires the specification of the covariance matrix of $\hat{\beta}$ as well as matrix $K$ used for the quadratic test static. Since in the *Full* method, the estimator used is $\hat{\beta}_o$ compared to $\hat{\beta}_{mc}$ one needs to show how matrix $K$ is effected. Define $\eta^* = 0 \ \forall i \neq j$, and $\eta_i^* = 1$, the Tag-SNP can be written as $\eta^*\hat{\beta}_m$.

That is the testing procedure is $\hat{\beta}_m'\eta^*\eta^{*'}\hat{\beta}_m > t$, where $t$ is some threshold. Therefore, for the *Full* procedure we can write the testing as,

$$\hat{\beta}_o'\hat{\Sigma}_o\eta^*\eta^{*'}\hat{\Sigma}_o\hat{\beta}_o > t \tag{17}$$

and for the Empirical corrected procedure,

$$\hat{\beta}'_{mc}\hat{\Sigma}_r\eta^*\eta^{*'}\hat{\Sigma}_r\hat{\beta}_{mc} > t. \tag{18}$$

Thus, both method have the same selection method, but different estimates of the coefficients as well as covariance matrices.

# S4    R-Package

[R-package for ECCCM routine:] R-package ECCCM containing code to obtain the variance inflation caused by using a reference panel for marginal linear regression transformation. The package can be found at https://github.com/tfrostig/ECCCM.



\end{document}